\documentclass{gGAF2e}

\graphicspath{{./fig/}{./png/}}
\usepackage{color}

\newcommand{\alphaem}{\ensuremath{\alpha_{\rm em}}}
\newcommand{\s}{\,{\rm s}}
\newcommand{\g}{\,{\rm g}}

\newcommand{\MeV}{\,{\rm MeV}}
\newcommand{\GeV}{\,{\rm GeV}}
\newcommand{\erg}{\,{\rm erg}}
\def\kB{k_{\rm B}}

\newcommand{\G}{\,{\rm G}}
\newcommand{\K}{\,{\rm K}}

\newcommand{\cm}{\,{\rm cm}}

\newcommand{\nab}{\bm{\nabla}}

\newcommand{\BB}{\bm{B}}

\newcommand{\JJ}{\bm{J}}
\newcommand{\UU}{\bm{U}}

\newcommand{\uu}{\bm{u}}

\newcommand{\SSSS}{\mbox{\boldmath ${\sf S}$} {}}
\newcommand{\meanrho}{\overline{\rho}}

{}
{}
{}

{}
{}
{}
{}
{}
{}
\newcommand{\meanAB}{\overline{\mbox{\boldmath $A\cdot B$}}{}}{}
{}
{}
\newcommand{\meanBB}{\overline{\mbox{\boldmath $B$}}{}}{}
{}
{}
{}
{}
{}
{}
{}
\newcommand{\meanJJ}{\overline{\mbox{\boldmath $J$}}{}}{}

{}
{}
{}

\newcommand{\meanmu}{\overline{\mu}}

\newcommand{\meanv}{\overline{v}}
\newcommand{\EQA}{\begin{eqnarray}}
\newcommand{\ENA}{\end{eqnarray}}
\newcommand{\eq}[1]{(\ref{#1})}

\renewcommand{\Gamma}{{\varGamma}}
\renewcommand{\Phi}{{\varPhi}}
\renewcommand{\Upsilon}{{\varUpsilon}}
\newcommand{\UpsilonTimeAve}{\langle\varUpsilon\rangle_t}

\newcommand{\prd}{Phys.\ Rev. D}
\newcommand{\pre}{Phys.\ Rev. E}
\newcommand{\apj}{Astrophys.\ J.}
\newcommand{\apjl}{Astrophys.\ J.\ Lett.}
\newcommand{\physrep}{Phys.\ Rep.}
\newcommand{\aap}{Astro.\ and Astrophys.}

\newcommand{\bra}[1]{\langle #1\rangle}

\renewcommand{\max}{{\rm max}}

\def\half{{\textstyle{1\over2}}}
\def\onethird{{\textstyle{1\over3}}}

\def\Pm{{\rm Pr}_{_\mathrm{M}}}
\def\Rm{{\rm Re}_{_\mathrm{M}}}

\def\cs{c_{\rm s}}

\begin{document}

\jvol{00} \jnum{00} \jyear{2012} 

\markboth{J.~Schober et al.}{GEOPHYSICAL AND ASTROPHYSICAL FLUID DYNAMICS}

\title{Energetics of turbulence generated by chiral MHD dynamos}

\author{
J. SCHOBER${^1}$$^{\ast}$\thanks{$^\ast$Corresponding author. Email: jennifer.schober@epfl.ch},
A. BRANDENBURG${^{2,3,4}}$,
I. ROGACHEVSKII${^{5,2}}$,
\& N. KLEEORIN${^{5,2}}$\\
\vspace{6pt}  ${^1}$ Laboratoire d'Astrophysique, EPFL, CH-1290 Sauverny, Switzerland\\
\vspace{6pt}  ${^2}$ Nordita, KTH Royal Institute of Technology
 and Stockholm University, Roslagstullsbacken 23,
 10691 Stockholm, Sweden \\
\vspace{6pt}  $^3$Laboratory for Atmospheric and Space Physics,
              JILA and Department of Astrophysical and Planetary Sciences,
              University of Colorado, Boulder, CO 80303, USA\\
\vspace{6pt}  $^4$Department of Astronomy,
    Stockholm University, SE-10691 Stockholm, Sweden\\
\vspace{6pt}  $^5$Department of Mechanical Engineering,
 Ben-Gurion University of the Negev, P.O. Box 653, Beer-Sheva
 84105, Israel
}

\vspace{6pt}\received{\today,~ $ $Revision: 1.186 $ $}

\maketitle

\begin{abstract}
An asymmetry in the number density of left- and right-handed fermions
is known to give rise to a new term in the induction equation
that can result in a dynamo instability.
At high temperatures, when a chiral asymmetry can survive for long enough,
this chiral dynamo instability
can amplify magnetic fields efficiently, which in turn drive
turbulence via the Lorentz force.
While it has been demonstrated in numerical simulations that
this chiral magnetically driven turbulence exists and
strongly affects the dynamics of the magnetic field,
the details of this process remain unclear.
The goal of this paper is to
analyse the energetics of chiral magnetically driven turbulence
and its effect on the generation and dynamics of magnetic field
using direct numerical simulations.
We study these effects for different initial conditions,
including a variation of the initial chiral chemical potential and the
magnetic Prandtl number, $\Pm$.
In particular, we determine
the ratio of kinetic to magnetic energy, $\Upsilon$,
in chiral magnetically driven turbulence.
Within the parameter space explored in this study, $\Upsilon$ reaches a value of
approximately $0.064$--$0.074$---independently of the initial chiral
asymmetry and for $\Pm=1$.
Our simulations suggest, that $\Upsilon$ decreases as a power law when
increasing $\Pm$ by decreasing the viscosity.
While the exact scaling depends on the details of the fitting criteria and the
Reynolds number regime, an approximate result of $\Upsilon(\Pm)=0.1\ \Pm^{-0.4}$
is reported.
Using the findings from our numerical simulations,
we analyse the energetics of
chiral magnetically driven turbulence in the early Universe.
\begin{keywords}
Relativistic magnetohydrodynamics (MHD); Chiral MHD dynamos; Turbulence; Early Universe
\end{keywords}

\end{abstract}

\section{Introduction}

Turbulence and magnetic fields are closely connected in many
geophysical and astrophysical flows:
Magnetohydrodynamic (MHD) dynamos are often related to
turbulence, so, for example, in the cases of the small-scale
\citep{Kazantsev1968,KulsrudAnderson1992} and large-scale
dynamos, especially those driven by
helical turbulent motions causing the $\alpha$ effect \citep{P55,SKR66}.
On the other hand, the Lorentz force resulting from magnetic fields can
drive turbulent motions.
How much magnetic energy can be converted into kinetic energy depends on
various properties of the plasma,
characterised by the fluid and magnetic Reynolds numbers,
and the structure of the magnetic field.
Hence, turbulence is a key ingredient for understanding the origin and
evolution of cosmic magnetic fields.

Observational constraints on the lower limits on the strength of intergalactic
magnetic fields \citep{NV10,DCRFCL11} challenge theoretical scenarios like the ones including
the turbulent dynamo.
A theory explaining these possible remains of primordial
fields includes the generation of seed fields
on small spatial scales, below the co-moving Hubble radius of the early Universe,
and a subsequent cascade to larger scales in decaying MHD turbulence either with
magnetic helicity \citep{BEO96,BM99PhRvL,FC2000,KTBN13,BKMPTV17} or
without \citep{Zra14,BKT15}.
Cosmological seed fields, however, are
a highly debated topic in modern cosmology; see e.g.\
\citet{GrassoRubinstein2001,KZ08,Subramanian16}.
In addition to various generation mechanisms suggested in the literature,
seed fields have recently been connected to a microphysical effect that is
related to the two opposite handedness of fermions.
In the presence of an external magnetic field, the momenta of fermions
align along the field lines according to their spin:
right-handed fermions are accelerated along the field lines, while left-handed ones are accelerated
in the opposite direction.
Collisions between particles lead to a constant flow along the field
lines, with the direction depending on the handedness.
Consequently, an asymmetry in the number density of left- and right-handed
charged particles leads to a net current along the magnetic field. 
This effect is
called the \textit{chiral magnetic anomaly}
\citep{Vilenkin:80a,RW85,Tsokos:85,AlekseevEtAl1998,Frohlich:2000en,
Frohlich:2002fg,Kharzeev:07,Fukushima:08,Son:2009tf}
and the resulting current can lead to a magnetic dynamo instability \citep{JS97}.
Especially the studies of the chiral inverse magnetic cascade and
the evolution of a non-uniform chiral chemical potential
by \citet{BFR12,BFR15} who found that a chiral asymmetry can, in principle,
survive down to energies of the order of $10\MeV$ ($\approx 10^{11}\K$), made
this effect an interesting candidate for cosmological applications.

Recently, a systematic analytical study of the system of chiral MHD
equations, including the back-reaction of the magnetic field on the
chiral chemical potential, and the coupling to the plasma velocity field has
been performed by \citet{REtAl17}.
High-resolution numerical simulations, presented in \citet{Schober2017},
confirm results from mean-field theory,
in particular the existence of a new chiral $\alpha$ effect
that is not related to the kinetic helicity, the so-called $\alpha_\mu$ effect.
Spectral properties of chiral MHD turbulence have been analysed in
\citet{BSRKBFRK17}.
A key result from these direct numerical simulations (DNS) is
that turbulence can be magnetically driven by the Lorentz force
due to a small-scale chiral dynamo instability.
In particular, a new three-stage-scenario of the magnetic field
evolution has been found in \citet{Schober2017}.
The small-scale chiral dynamo instability is followed by a phase in which
magnetically produced turbulence triggers a large-scale dynamo
instability, which eventually saturates due to the decrease of the chiral
chemical potential.

In this paper we extend the work of \citet{Schober2017}
to analyse the energetics of chiral magnetically driven turbulence.
We explore
different initial values
of the chiral chemical potential to find
the dependence of the ratio of the kinetic energy over the
magnetic energy on the initial chiral asymmetry and the magnetic Prandtl number,
$\Pm\equiv \nu/\eta$, where $\nu$ is the kinematic viscosity and $\eta$ is the magnetic
diffusivity.
We also determine energy transfer rates
between the different energy reservoirs and
obtain the dependence of the ratio of kinetic to magnetic energy dissipation rates
on the magnetic Prandtl number.

The paper is structured as follows: In section~\ref{sec_ChiralMHD}, we outline
the chiral MHD equations, the growth rates of their instabilities, and the
saturation magnetic fields expected from the conservation law in chiral MHD.
In this section, we also discuss the different stages of magnetic field evolution
and the production of chiral magnetically driven turbulence.
The setup of our numerical simulations is described in section~\ref{sec_DNS}
and compared with those presented in \citet{Schober2017}.
We discuss here also the results of the direct numerical simulations related
to the dynamics of the velocity and magnetic fields.
In section~\ref{subsec_urmsBrms}, we analyse the ratio
of kinetic over magnetic energy for different dynamo growth rates and
different magnetic Prandtl numbers.
Additionally, the transfer of energy
from the chiral chemical potential, via magnetic energy,
to turbulent kinetic energy
is studied by determining the energy production and dissipation rates.
In section~\ref{sec_EU}, we estimate the magnetic Prandtl and Reynolds numbers
in the relativistic plasma of the early Universe and apply our results on the
magnetic Prandtl number dependence.

\section{Chiral MHD}
\label{sec_ChiralMHD}

\subsection{Governing equations}

We begin by reviewing the basic equations of chiral MHD, as derived
by \citet{REtAl17}.
We consider the case of very low microscopic magnetic diffusivity, $\eta$,
which is the relevant regime for astrophysical applications.
The chiral asymmetry is described by the chiral chemical potential,
\begin{eqnarray}
 \mu_5=6\,(n_{\rm L}-n_{\rm R})\,{(\hbar c)^3\over(\kB T)^2},
\end{eqnarray}
which is proportional to the difference in the number densities of left- and
right-chiral fermions, $n_\mathrm{L}$ and $n_\mathrm{R}$, respectively.
Here, $T$ is the temperature, $\kB$ is the Boltzmann constant,
$c$ is the speed of light, and $\hbar$ is the reduced Planck constant.
In an external magnetic field, $\mu_5$ gives rise to a current
due to the chiral magnetic effect (CME)
\begin{eqnarray}
  \label{eq_CME}
   \JJ_{\rm CME} = \frac{\alphaem}{\pi \hbar} \mu_5 \BB ,
\end{eqnarray}
where $\alphaem \approx 1/137$ is the fine structure constant.
This quantum relativistic effect, described by the standard model of particle physics,
results in an additional term in the Maxwell equations.
Based on these modified Maxwell equations,
\citet{BFR15} and \citet{REtAl17} derived the following set of chiral MHD equations:
\begin{eqnarray}
  \frac{\partial \BB}{\partial t} &=& \nab   \times   \left[{\UU}  \times   {\BB}
  - \eta \, \left(\nab   \times   {\BB}
  - \mu {\BB} \right) \right] ,
\label{ind-DNS}\\
  \rho{D \UU \over D t}&=& (\nab   \times   {\BB})  \times   \BB
  -\nab  p + \nab  {\bm \cdot} (2\nu \rho \SSSS) ,
\label{UU-DNS}\\
  \frac{D \rho}{D t} &=& - \rho \, \nab  \cdot \UU ,
\label{rho-DNS}\\
  \frac{D \mu}{D t} &=& D_5 \, \Delta \mu
  + \lambda \, \eta \, \left[{\BB} {\bm \cdot} (\nab   \times   {\BB})
  - \mu {\BB}^2\right]
\label{mu-DNS}
\end{eqnarray}
where
$\UU$ is the fluid velocity,
the magnetic field $\BB$ is normalised such that the magnetic energy
density is $\BB^2/2$ (so the magnetic field in Gauss is $\sqrt{4\pi}\,\BB$), and
$D/D t = \partial/\partial t + \UU \cdot \nab$ is the
advective derivative.
Further, a normalisation of $\mu_5$ is used such that
$\mu = (4 \alphaem /\hbar c) \mu_5$ and the chiral feedback parameter $\lambda$ 
has been introduced, which characterises the strength of the
back-reaction from the electromagnetic
field on the evolution of $\mu$.
For hot plasmas, when $\kB T \gg \max(|\mu_L|,|\mu_R|$),
it is given by \citep{BFR15}
\begin{eqnarray}
  \lambda=3 \hbar c \left({8 \alphaem \over \kB T} \right)^2.
\label{eq_lambda}
\end{eqnarray}
In equations~(\ref{ind-DNS})--(\ref{mu-DNS}),
$D_5$ is a chiral diffusion coefficient, $p$ is the fluid pressure,
${\sf S}_{ij}=\half(U_{i,j}+U_{j,i})-\onethird\delta_{ij} {\bm \nabla}
{\bm \cdot} \UU$
are the components of the trace-free strain tensor, where commas denote partial
spatial differentiation.
For an isothermal equation of state, the pressure $p$ is related
to the fluid density $\rho$
via $p=c_{\rm s}^2\rho$, where $c_{\rm s}$ is the isothermal sound speed.
Flipping reactions between right- and left-handed states of fermions have been
neglected in equations~(\ref{ind-DNS})--(\ref{mu-DNS}).
For an overview of the parameters and characteristic scales governing chiral 
MHD, we refer to table~\ref{tab_chiralMHDunits}.

\textbf{
\begin{table}[t!]
  \centering
  \caption{Overview of the parameters in chiral MHD. Units are given in CGS with
the corresponding Natural Unit in brackets.}
  \label{tab_chiralMHDunits}
  \begin{tabular}{lllll}
\hline
\hline
  \emph{Parameter} 	 			& \emph{Symbol} 	& \emph{Unit}      								& \emph{Definition}    \\
\hline
\emph{chiral MHD parameters:}   \\
  chiral chemical potential			& $\mu_5$     		& erg~							[eV]			&  $6\,(n_{\rm L}-n_{\rm R})\,{(\hbar c)^3/(\kB T)^2}$	 \\
  (normalised) "				& $\mu$      		& $\mathrm{cm}^{-1}$ 					[eV]			&  $4 \alphaem/(\hbar c) \mu_5$  \\
  initial value of $\mu$			& $\mu_0$      		& $\mathrm{cm}^{-1}$ 					[eV]			&    \\
  chiral velocity				& $v_{\mu}$		& -										&  $\eta\mu_0$  \\
  chiral Mach number				& ${\rm Ma}_\mu$	& -										&  $v_\mu/c_\mathrm{s}$  \\
  chiral non-linearity parameter		& $\lambda$		& $\mathrm{s}^{2}\mathrm{g}^{-1}\mathrm{cm}^{-1}$ 	[$\mathrm{eV}^{-2}$]	&  $3 \hbar c \left(8 \alphaem/( \kB T) \right)^2$	\\
  (non-dimensional) "				& $\lambda_\mu$	& -	 										&  $\lambda \eta^2 \meanrho$ \\
  chiral diffusivity   				& $D_5$			& $\mathrm{cm}^{2}\mathrm{s}^{-1}$			[$\mathrm{eV}^{-1}$]	& 	\\
  chiral diffusion rate  			& $\epsilon_\mu$	& $\mathrm{erg}~\mathrm{s}^{-1}$		 	[$\mathrm{eV}^{2}$]	&  $\langle D_5 \nabla^2\mu\rangle$ 	\\
\hline
\emph{classical MHD parameters:}   \\
  magnetic diffusivity   			& $\eta$		& $\mathrm{cm}^{2}\mathrm{s}^{-1}$			[$\mathrm{eV}^{-1}$]	& 	\\
  kinematic viscosity  					& $\nu$			& $\mathrm{cm}^{2}\mathrm{s}^{-1}$			[$\mathrm{eV}^{-1}$]	& 	\\
  turbulent diffusivity   			& $\eta_{_{T}}$		& $\mathrm{cm}^{2}\mathrm{s}^{-1}$			[$\mathrm{eV}^{-1}$]	& $\approx u_\mathrm{rms}/(3k_\mathrm{f})$	\\
  kinetic energy dissipation rate			& $\epsilon_\mathrm{K}$  & $\mathrm{erg}~\mathrm{s}^{-1}$			[$\mathrm{eV}^{2}$]	& $\langle 2 \nu \rho \SSSS^2\rangle$	\\
  magnetic energy diffusion rate		& $\epsilon_\mathrm{M}$  & $\mathrm{erg}~\mathrm{s}^{-1}$			[$\mathrm{eV}^{2}$]	&  $\langle \eta \JJ^2\rangle$	\\
  " of mean magnetic field		& $\tilde{\epsilon}_\mathrm{M}$  & $\mathrm{erg}~\mathrm{s}^{-1}$			[$\mathrm{eV}^{2}$]	&  $\langle (\eta+\eta_{_{T}}) \meanJJ^2\rangle$	\\
\hline
\emph{characteristic wavenumbers:} \\
  small-scale chiral instability		& $k_\mu$		& $\mathrm{cm}^{-1}$ 					[eV]			&  $\mu_0/2$ \\
  $\alpha_\mu$ instability			& $k_\alpha$		& $\mathrm{cm}^{-1}$ 					[eV]			&  $|\meanv_\mu + \alpha_\mu| / (2 \eta+ \, 2 \eta_{_{T}})$ \\
  saturation			      		& $k_\lambda$		& $\mathrm{cm}^{-1}$ 					[eV]			&  $\sqrt{\meanrho\lambda C_\mu/C_\lambda}\,\eta\mu_0$ \\
  box size					& $k_1$			& $\mathrm{cm}^{-1}$ 					[eV]			&  $1/L$ \\
\hline
\emph{characteristic growth rates:} \\
  small-scale chiral instability		& $\gamma_\mu$		& $\mathrm{s}^{-1}$ 					[eV]			&  $\eta\mu_0^2/4$ \\
  $\alpha_\mu$ instability			& $\gamma_\alpha$	& $\mathrm{s}^{-1}$ 					[eV]			&  $(\meanv_\mu + \alpha_\mu)^2/(4 (\eta+ \, \eta_{_{T}})) $ \\
\hline
\emph{characteristic field strengths:} \\
  initial value 				& $B_0$			& G							[$\mathrm{eV}^{2}$]	&   \\
  transition: laminar to turbulent 		& $B_\mathrm{rms}^{1\to2}$	& G						[$\mathrm{eV}^{2}$]	&  $\approx  \left(C_\mu\meanrho/2\right)^{1/2} \mu_0 \eta$ \\
  dynamo saturation				& $B_\mathrm{sat}$	& G							[$\mathrm{eV}^{2}$]	&  $\approx \left(\meanrho \eta^2 C_\mu C_\lambda/\lambda\right)^{1/4} \mu_0$ \\
\hline
\emph{dimension less parameters:} \\
  energy ratio					& $\Upsilon$		& -										&  $\rho u_\mathrm{rms}^2/B_\mathrm{rms}^2$ \\
  ratio of production rates			& $\Phi$		& -										&  $\langle \UU \cdot (\JJ\times\BB)\rangle/(|v_\mu \BB \cdot \JJ|)$ \\
\hline
\hline
  \end{tabular}
\end{table}
}

The system of equations is determined by several non-dimensional parameters.
In terms of chiral MHD dynamos, the most relevant ones are the chiral
Mach number
\begin{eqnarray}
  {\rm Ma}_\mu = \frac{\eta\mu_0}{c_\mathrm{s}} \equiv \frac{v_{\mu}}
  {c_\mathrm{s}},
\label{Ma_mu_def}
\end{eqnarray}
where $\mu_0$ is the initial value of $\mu$,
and the dimensionless chiral nonlinearity parameter:
\begin{eqnarray}
  \lambda_\mu = \lambda \eta^2 \meanrho.
\label{eq_lambdamu}
\end{eqnarray}
The parameter ${\rm Ma}_\mu$ measures the relevance
of the chiral term in the induction
equation~(\ref{ind-DNS}) and determines
the growth rate of the small-scale chiral dynamo instability.
The nonlinear back reaction
of the magnetic field on the chiral chemical potential $\mu$
is characterised by $\lambda_\mu$,
which affects the strength of the saturation magnetic field and
the strength of the magnetically driven turbulence.
In this paper we consider only cases with $\lambda_\mu \ll 1$, i.e., when turbulence
is produced efficiently due to strong magnetic fields
generated by the small-scale chiral dynamo instability.
The turbulent cascade properties have previously been studied by
\citet{BSRKBFRK17} in the range $2\times10^{-6}\leq\lambda_\mu \leq200$,
\citet{Schober2017} in the range $10^{-9}\leq\lambda_\mu \leq10^{-5}$.

\begin{figure}
\begin{center}
  \includegraphics[width=0.8\textwidth]{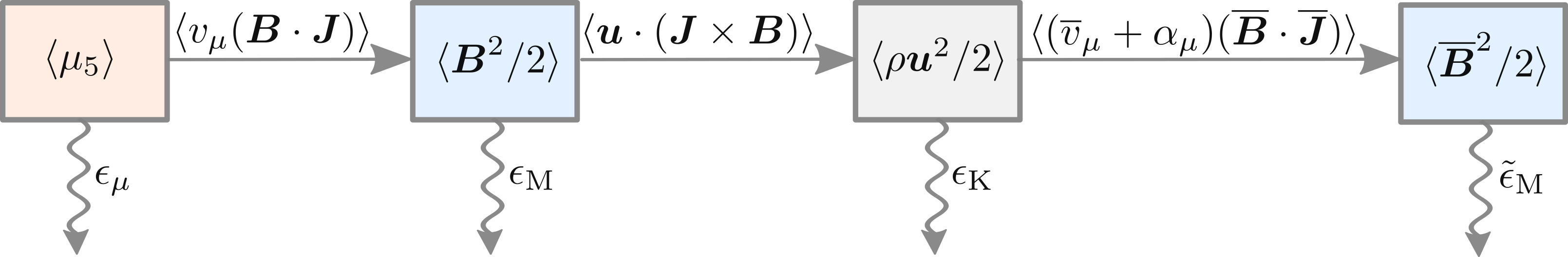}
\caption{
Illustration of energy transfer
from the chiral chemical potential
$\langle\mu_5\rangle$, to magnetic energy $\langle\BB^2/2\rangle$,
further to turbulent kinetic energy,
$\langle\rho \uu^2/2\rangle$ and finally to energy of the large-scale magnetic field, $\meanBB^2/2$.
Losses via microscopic magnetic diffusion and kinematic viscosity
are indicated by curled arrows. (colour online)
}
\label{fig_energyflow}
\end{center}
\end{figure}

The illustration in figure~\ref{fig_energyflow} shows how energy is converted
from a chiral chemical potential, to magnetic energy,
further to turbulent kinetic energy,
and finally to energy of the large-scale
magnetic field.
The relevant transport terms are indicated in the sketch
together with the diffusion terms, $\epsilon_\mu \equiv \langle D_5 \nabla^2
\mu\rangle$, $\epsilon_\mathrm{M} \equiv \langle \eta \JJ^2\rangle$,
$\epsilon_\mathrm{K} \equiv \langle 2 \nu \rho \SSSS^2\rangle$, and
$\tilde{\epsilon}_\mathrm{M} \equiv \langle (\eta+\eta_{_{T}}) \meanJJ^2\rangle$.
Here, $\JJ$ and $\meanJJ$ are the total and mean values of the electric
current, respectively, and $\eta_{_{T}}$ is the turbulent
magnetic diffusivity. 
The latter is defined as $\eta_{_{T}}=u_\mathrm{rms}/(3 k_\mathrm{f})$, where
$u_\mathrm{rms}$ is the rms velocity and $k_\mathrm{f}$ integral length scale
of turbulence.

\subsection{Analogy with the $\alpha$ effect in mean-field MHD}

Readers familiar with mean-field MHD \citep[see][mfMHD]{M78,KR80}
will have readily noticed the analogy between
$v_\mu=\mu \, \eta$ in chiral MHD
and the kinetic part of the $\alpha$ effect, $\alpha_{\rm K}$,
in mfMHD.
For $\eta\to 0$, the analogy goes further in that even the evolution
equation~\eq{mu-DNS} for $\mu$ corresponds to an analogous one
for the magnetic part of the $\alpha$ effect, $\alpha_{\rm M}$,
proportional to the magnetic helicity,
in what is known as the dynamical quenching formalism \citep{KR82,KRR95}.

Exploiting the analogy between chiral MHD and dynamical quenching
can be beneficial in two ways.
First, there is a considerable body of work on dynamical quenching that
can improve our intuition in chiral MHD \citep[e.g.][]{KMRS2000,BB02}.
Second, numerical approaches have been developed for dynamical quenching
that can directly be utilised in chiral MHD.
The purpose of this section is to elaborate on this analogy, which was
never mentioned before.
Readers unfamiliar with dynamical quenching may skip forward to
section~\ref{sec_overview}.

In chiral MHD, dynamical quenching means that the total chirality,
i.e., the sum of magnetic helicity and fermion chirality, is conserved.
In mfMHD, in the absence of magnetic helicity fluxes,
it implies that the total magnetic helicity is conserved,
i.e., the sum of the magnetic helicity of the mean-field and that
of the fluctuating field, the latter of which constitutes an
additional time-dependent contribution to the $\alpha$ effect.
The other contribution to the $\alpha$ effect
in mfMHD, $\alpha_{\rm K}$, is proportional
to the kinetic helicity, which was here assumed to be constant in time, so we can write
\citep[see equation 18 of][]{BB02}
\begin{equation}
  \frac{\partial \alpha}{\partial t} =
  \lambda_{\rm mfMHD} \, \eta \left[\eta_{_{T}} {\meanBB} {\bm \cdot} (\nab   \times   {\meanBB})
  - \alpha {\meanBB}^2\right]
  -\Gamma_{\rm mfMHD}\, (\alpha-\alpha_{\rm K}),
\end{equation}
where
$\alpha=\alpha_{\rm K}+\alpha_{\rm M}$.
In mfMHD, the coupling coefficient is given by
$\lambda_{\rm mfMHD}=2 \eta_{_{T}} k_{\rm f}^2/(\eta B_{\rm eq}^2)$ and
$\Gamma_{\rm mfMHD}
=2\eta k_{\rm f}^2$, where
$k_{\rm f}$ is the wavenumber of the energy-carrying eddies and
$B_{\rm eq}$ is the equipartition field strength.

The applications of chiral MHD carry over to decaying MHD turbulence with
finite initial large-scale or small-scale magnetic helicity \citep{KBJ11}.
During the decay, some of the magnetic helicity is transferred between
the large- and small-scale fields, which leads to a change in the $\alpha$
effect that in turn results in a slow-down of the decay.

\subsection{Review of the three stages of the magnetic field evolution}
\label{sec_overview}

Recent simulations by \citet{Schober2017} have demonstrated the existence
of three distinct stages characterising
the growth and saturation of the magnetic field in different kinds of
chiral dynamos: \\
\textbf{Phase 1}:
a laminar phase of small-scale chiral dynamo instability; \\
\textbf{Phase 2}:
a large-scale dynamo instability,
caused by chiral magnetically produced turbulence; \\
\textbf{Phase 3}:
termination of growth of the large-scale magnetic field
and reduction of $\mu$ according to the conservation law in chiral MHD. \\
With no further energy input, dynamo saturation
is followed by decaying helical MHD turbulence, where the magnetic field
decreases with time as a power law like $|\BB|\sim t^{-1/3}$
\citep{BM99PhRvL, KTBN13}.

\begin{figure}
\begin{center}
  \includegraphics[width=0.49\textwidth]{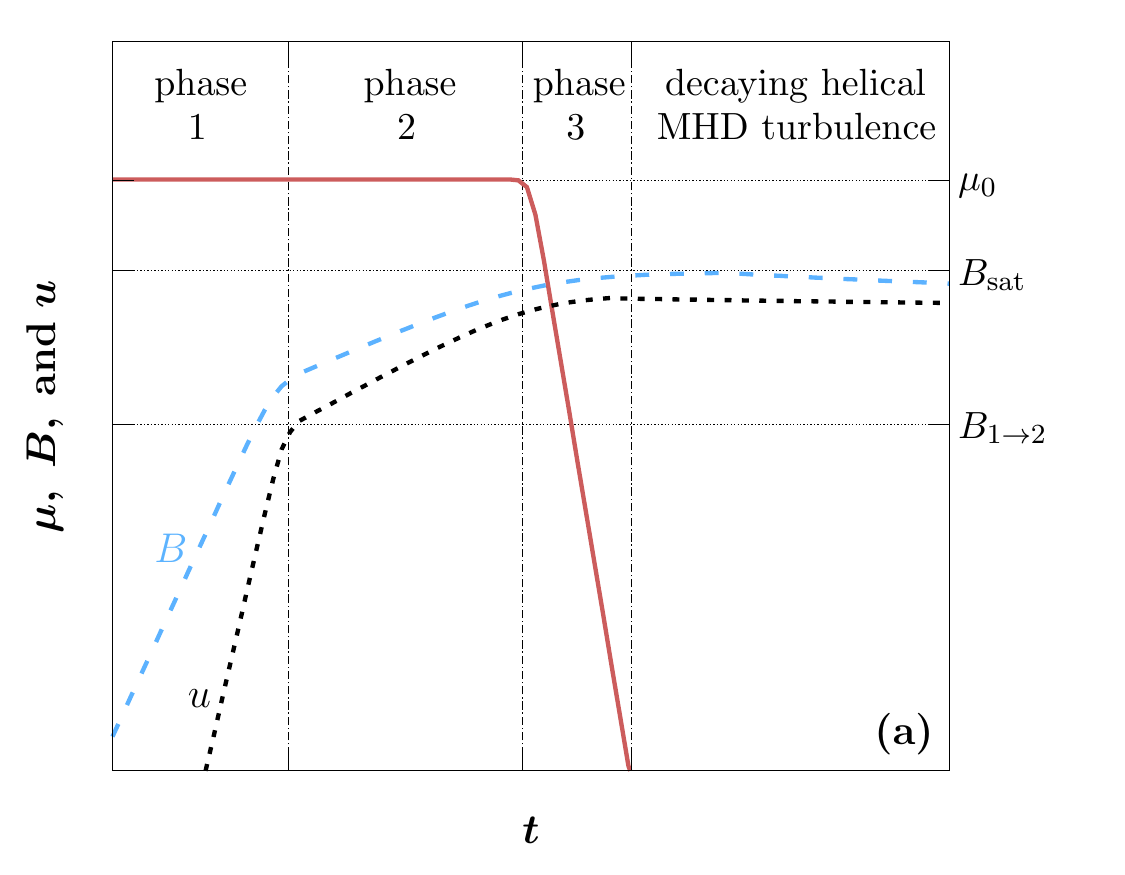}
  \includegraphics[width=0.49\textwidth]{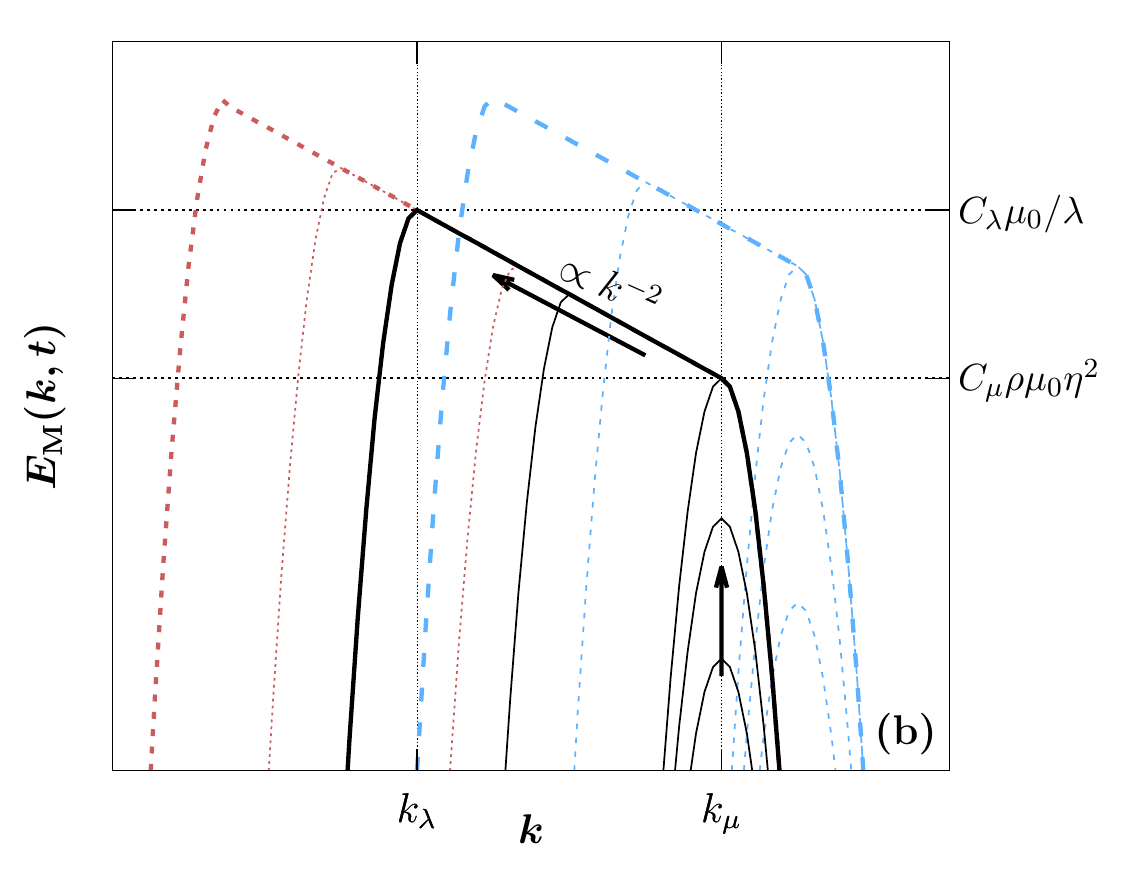}
\caption{
Schematic overview of the evolution of a plasma in an infinite box
with an initially weak magnetic field in the presence of a chiral chemical potential.
For a detailed discussion see section~\ref{sec_overview}.
(a):
The time evolution of $\mu$ (red solid line), $\uu$ (black dotted
line), and $\BB$ (blue dashed line).
The values of $\mu_0$ and $B_\mathrm{sat}$ are indicated as
horizontal dotted lines.
The transitions between individual phases of the evolution are marked by vertical
dashed-dotted lines.
(b):
Evolution of magnetic energy spectra with inverse transfer from $k_\mu$ to $k_\lambda$.
The blue lines correspond to a larger $\mu_0$ in comparison with the case
shown by the black lines (but for the same $\lambda$), while
the red lines correspond to a smaller $\lambda$ in comparison with the case
indicated by the black lines (but for the same $\mu_0$). (colour online)
}%
\label{fig_sketch}
\end{center}
\end{figure}

In this section, the dynamics of chiral magnetically driven turbulence is
shortly reviewed for the case of a chiral
plasma in an infinite domain (see figure~\ref{fig_sketch}).
In section~\ref{sec_finiteboxeffects}, we will discuss the potential discrepancies
in this picture arising from the effect of a finite
computational domain
(see figure~\ref{fig_sketch_box}).

\subsubsection{Amplification of magnetic fields by chiral dynamos}

In \textbf{phase 1}, the velocity field is negligible
and a small-scale laminar dynamo operates.
The growth rate found from the linearised equation~(\ref{ind-DNS}) has a
maximum value of \citep{JS97}
\begin{eqnarray}
  \gamma_\mu = \frac{v_\mu^2}{4 \eta}
\label{gamma-max}
\end{eqnarray}
being attained at
\begin{equation}
  k_\mu =\frac{|\mu_0|}{2}.
\label{eq_kmax}
\end{equation}
While the magnetic field grows with a rate $\gamma_\mu$,
turbulence is driven by the Lorentz force with the rms
velocity increasing at a rate of approximately $2 \gamma_\mu$.

In \textbf{phase 2},
the turbulent velocity has become so large that it
affects the evolution of the magnetic field.
It has been shown by \citet{BSRKBFRK17} that the peak of the magnetic energy
spectrum reaches a value of
\begin{equation}
   E_\mathrm{M}^{1\to2} = C_\mu\meanrho \mu_0 \eta^2
\end{equation}
with $C_\mu\approx 16$ at the transition from phase 1 to phase 2.
This moment coincides with the beginning of the inverse transfer, when the
$k^{-2}$ spectrum starts to build up, i.e.\ when the peak of the magnetic energy
spectrum moves from $k_\mu$ to smaller wavenumbers.
The corresponding transition field strength can be estimated as
\begin{equation}
   B_\mathrm{rms}^{1\to2} \approx  \left(E_\mathrm{M}^{1\to2} k_\mu\right)^{1/2}
                   \approx  \left(\frac{C_\mu\meanrho}{2}\right)^{1/2} \mu_0 \eta.
\end{equation}

At this stage, the chiral magnetically produced turbulence causes excitation
of a large-scale magnetic field by the chiral $\alpha_\mu$ effect.
This chiral large-scale dynamo, studied by \citet{REtAl17}, occurs at
the maximum growth rate
\begin{equation}
\gamma_\alpha
   = {(\meanv_\mu + \alpha_\mu)^2\over 4 (\eta+ \, \eta_{_{T}})}
   = {(\meanv_\mu + \alpha_\mu)^2\over 4 \eta \, (1 + \, \Rm/3)},
\label{gammamax_turb}
\end{equation}
where $\meanv_\mu \equiv \eta \meanmu$ and $\meanmu$ is the mean chiral chemical 
potential.
Despite the contribution from the chiral $\alpha_\mu$ effect, given by the term
$\alpha_\mu = - {2 \over 3} \meanv_\mu \ln \Rm$,
the overall growth rate is reduced as compared to the laminar chiral dynamo.
Here, $\Rm$ is the magnetic Reynolds number defined by
$\Rm=u_\mathrm{rms} /(\eta k_\mathrm{f}) = 3 \eta_{_{T}}/\eta$
The maximum growth rate of the chiral large-scale dynamo is attained at
the wavenumber
$k_\alpha = |\meanv_\mu + \alpha_\mu| / (2 \eta+ \, 2 \eta_{_{T}})$.

\subsubsection{Saturation of the chiral large-scale dynamo}

Saturation of the chiral large-scale dynamo, \textbf{phase 3},
is controlled by the conservation law following from 
equations~(\ref{ind-DNS})--(\ref{mu-DNS}), which
implies that the total chirality
\begin{equation}
  \frac\lambda 2 \meanAB + \bar\mu = \mu_0 = \mathrm{const},
\label{CL}
\end{equation}
is a conserved quantity; see \citet{REtAl17} for more details.
Here, $\meanAB$ is the spatially averaged value of
the magnetic helicity.
According to the conservation law (\ref{CL}), the magnetic field reaches
$B_\mathrm{sat} = \left(\mu_0/(\lambda\xi_\mathrm{M})\right)^{1/2}$,
where $\xi_\mathrm{M}$ is the correlation length of the magnetic field.

The magnetic energy spectrum $E_{\rm M}(k,t)$ in chiral MHD turbulence has been
studied in \citet{BSRKBFRK17}.
In particular, it was found that $E_{\rm M}$ is proportional to $k^{-2}$
between the wavenumber
\begin{equation}
   k_\lambda=\sqrt{\meanrho\lambda \frac{C_\mu}{C_\lambda}}\,\eta\mu_0,
\label{klambda}
\end{equation}
with $C_\mu\approx 16$, $C_\lambda\approx 1$, and $k_\mu$ given
by equation~(\ref{eq_kmax}).
We note that the only case considered here is $\lambda_\mu \ll 1$, which implies
$k_\lambda \ll k_\mu$.
Using dimensional arguments and numerical simulations,
\citet{BSRKBFRK17} found that for chiral magnetically driven turbulence,
the saturation magnetic energy spectrum $E_{\rm M}(k,t)$ obeys
\begin{equation}
   E_{\rm M}(k,t)=C_\mu\,\meanrho\mu_0^3\eta^2k^{-2}
\label{Cmu}
\end{equation}
in $k_\lambda<k<\mu_0$.
Here, $E_{\rm M}(k,t)$ is normalised such that
the mean magnetic energy density is
$\bra{\BB^2}/2=\int E_{\rm M}(k)\,\mathrm{d}k$.
It was also confirmed numerically by \cite{BSRKBFRK17}
that the magnetic energy spectrum $E_{\rm M}(k)$ is limited from above
by $C_\lambda \mu_0/\lambda$.
The magnetic field strength at dynamo saturation can be estimated as
\begin{equation}
   B_\mathrm{sat} \approx \left(E_{\rm M}(k_\lambda) k_\lambda\right)^{1/2}
                  =  \left(\frac{\meanrho \eta^2 C_\mu C_\lambda}{\lambda}\right)^{1/4} \mu_0.
\label{eq_Bsat}
\end{equation}

In figure~\ref{fig_sketch}(a) the effect of changing $\mu_0$
and $\lambda$ on the final energy spectrum is illustrated.
Here, intermediate spectra are shown as thin lines, while thick lines indicate
the magnetic energy spectrum at saturation with an inertial range where
$E_\mathrm{M} \propto k^{-2}$ between $k_\lambda$ and $k_\mu$.
The black lines present a case with a certain $\mu_\mathrm{black}$ and
$\lambda_\mathrm{black}$.
If $\lambda$ is decreased,
$k_\lambda$ also decreases, and the $k^{-2}$ spectrum spans
over a larger range of wavenumbers.
This is illustrated by the red lines, where
$\mu_\mathrm{red} = \mu_\mathrm{black}$ and
$\lambda_\mathrm{red} < \lambda_\mathrm{black}$.
In this case, the final magnetic field strength is higher than in the case
with $\mu_\mathrm{black}$ and $\lambda_\mathrm{black}$.
The same final field strength can, however, also be reached by
increasing $\mu_0$, as $B_\mathrm{sat}\propto \mu_0/\lambda^{1/4}$.
Schematic spectra illustrating the latter case are shown as blue curves in
figure~\ref{fig_sketch}, where $\mu_\mathrm{blue}>\mu_\mathrm{black}$ and
$\lambda_\mathrm{blue}=\lambda_\mathrm{black}$.

\subsection{Effects of the finite numerical domain}
\label{sec_finiteboxeffects}

\begin{figure}
\begin{center}
  \includegraphics[width=0.49\textwidth]{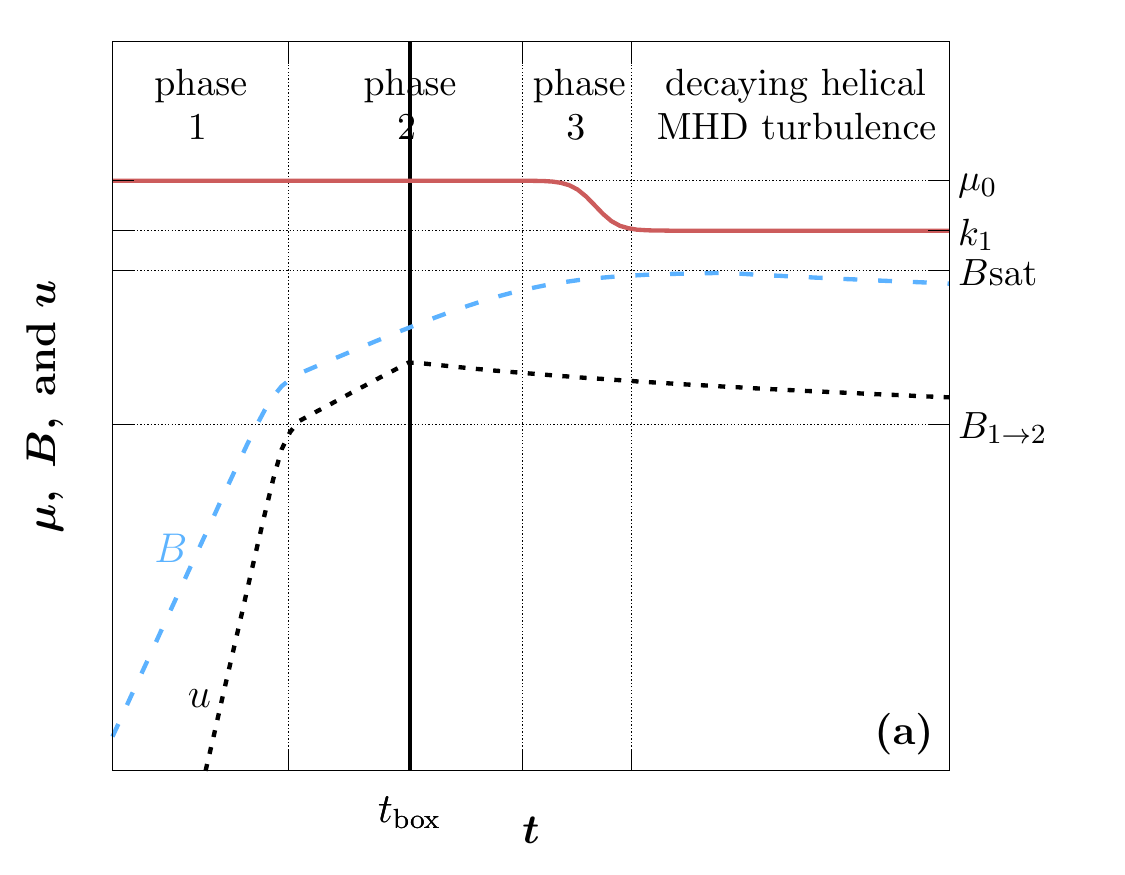}
  \includegraphics[width=0.49\textwidth]{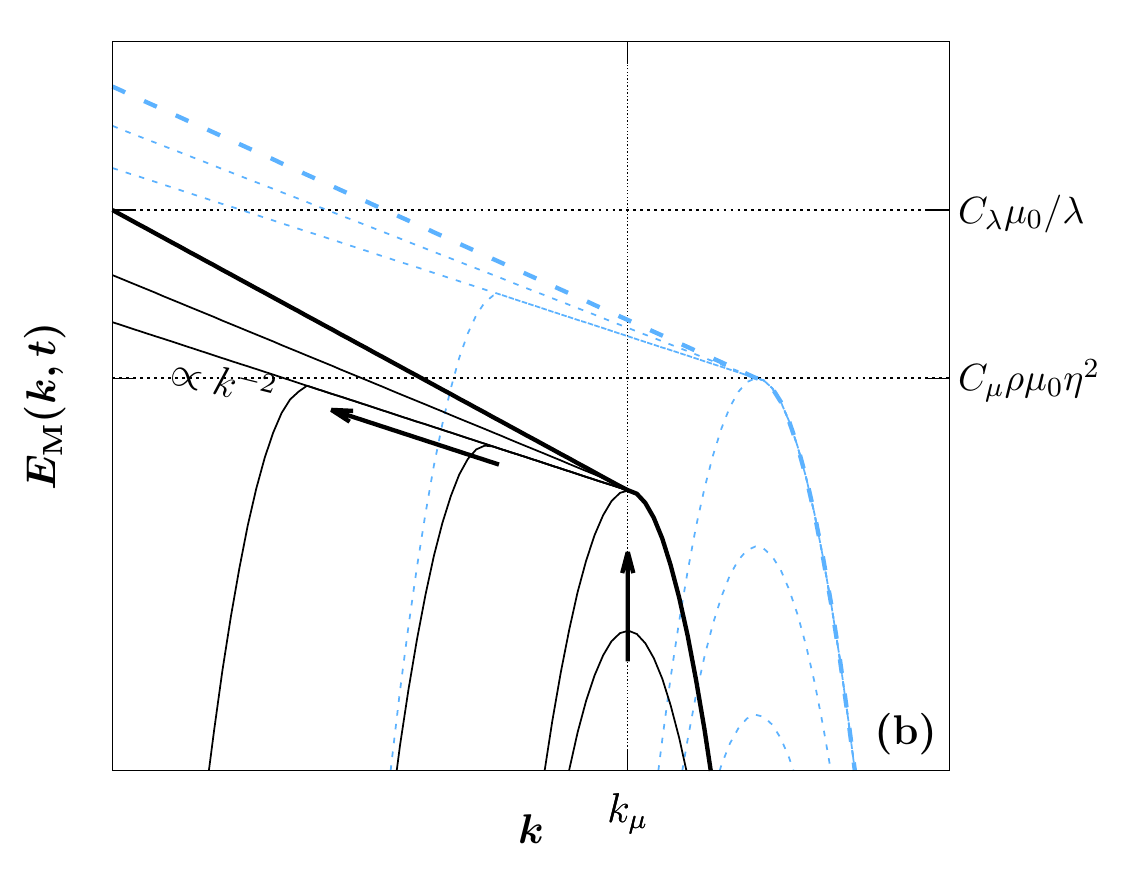}
\caption{Same as figure~\ref{fig_sketch} but including effects of a finite
box. (colour online)
}%
\label{fig_sketch_box}
\end{center}
\end{figure}

Due to a finite simulation domain, the evolution of the magnetic field
and the turbulent velocity is slightly modified in comparison to
the case discussed in section~\ref{sec_overview}.
The evolution of
the chiral chemical potential,
the magnetic field strength,
the rms velocity, and the time evolution of the magnetic energy spectra
in finite box simulations are illustrated in figure~\ref{fig_sketch_box}.

First of all, the chiral chemical potential,
does not vanish at dynamo saturation,
but reaches a finite value, which is equal to the minimum
wavenumber possible in the box, $k_1$.
The magnetic field reaches the saturation value
given by equation~(\ref{eq_Bsat}).
However, the evolution of the magnetic energy spectrum differs compared to
that anticipated for an infinite system.
In the laminar chiral dynamo phase, we expect an instability
at wavenumber $k_\mu$,
as predicted by theory; see the black curves in figure~\ref{fig_sketch_box}(b).
With the production of turbulence,
the peak of the magnetic energy spectrum moves to larger
spatial scales through inverse transfer.
As discussed above, we expect a scaling of the magnetic energy spectrum
proportional to $k^{-2}$; see equation~(\ref{Cmu}).
Once the peak reaches the size of the box, however, we observe a steepening
of the spectrum, as indicated in the schematic figure~\ref{fig_sketch_box},
if $k_\lambda<k_1$.
This steepening is caused by the growth of the magnetic field on the smallest
possible wavenumber, until the spectrum reaches its saturation value
$C_\lambda \mu_0/\lambda$.
For large values of $\mu_0$, the initial chiral dynamo instability
occurs at smaller spatial scales, i.e., at larger wavenumbers $k$,
and thus the $k^{-2}$ spectrum can extend over a
larger range; see the blue curves in
figure~\ref{fig_sketch_box}(b).
In this paper, we present also a run with $k_\lambda\approx k_1$, run~B, which
has a resolution of $1216^3$.
Large parameter scans, and cases with larger scale separation,
$k_\lambda \gg k_1$, are computationally too expensive.

\section{Chiral magnetically driven turbulence in direct numerical simulations}
\label{sec_DNS}

\subsection{Numerical setup}

We solve equations~(\ref{ind-DNS})--(\ref{mu-DNS})
in a three-dimensional
periodic domain of size $L^3 = (2\pi)^3$ with the
\textsc{Pencil Code}\footnote{\textit{http://pencil-code.nordita.org/}}.
This code is well suited for MHD studies; it employs a third-order
accurate time-stepping method of \cite{Wil80} and sixth-order explicit finite differences
in space \citep{BD02,Bra03}.
The smallest wavenumber covered in the numerical domain is $k_1 = 2\pi/L = 1$ and
the resolution is varied between $480^3$ and $1216^3$.
For comparison, we also show some simulations that have previously been presented
\citep{Schober2017}, but we now include additional runs for $\Pm\neq1$.
The sound speed in the simulations is set to $\cs = 1$ and the
mean fluid density to $\meanrho = 1$.
If not indicated otherwise, the magnetic Prandtl number is $1$,
i.e.\ the magnetic diffusivity equals the viscosity.
However, we do consider cases between $\Pm=0.5$ and $\Pm=10$, where the value of $\eta$
is fixed and $\nu$ changes.
No external forcing is applied to drive turbulence in these simulations, i.e.,
the velocity field is then driven entirely by magnetic fields.
All runs are initialised with a weak seed magnetic field in the form of
Gaussian noise, with constant $\mu$, and vanishing velocity.
The main input parameters of all simulations presented in this paper are 
summarised in table~\ref{table}.

\begin{table}
\centering
\caption{Overview of the input parameters of all simulations discussed in this paper.
We list the values of the non-dimensional chiral parameters, ${\rm Ma}_\mu$
and $\lambda_\mu$, see equations~(\ref{Ma_mu_def}) and (\ref{eq_lambdamu}),
respectively, as well as the characteristic wavenumbers, $k_\mu$
and $k_\lambda$, see equations~(\ref{eq_kmax}) and (\ref{klambda}),
normalized by the minimum wavenumber
corresponding to the finite numerical domain, $k_1$.
Runs~D, F, G and H have been performed
with different magnetic Prandtl numbers,
i.e., run~D05 with $\Pm=0.5$, run~D1 with $\Pm=1$,
up to run~D10 with $\Pm=10$.
Reference runs are highlighted by bold font.}
     \begin{tabular}{l|lllllll}
      \hline
      \hline
	simulation 	& resolution		& ${\rm Ma}_\mu$ 	& $\lambda_\mu$ 	& $(\mu_0/\lambda)^{1/2}$ 	& $k_\mu/k_1$ 	& $k_\lambda/k_1$	& $\Pm$\\	
      \hline
      	\textbf{Run A} 	& $\mathbf{576^3}$	& $\mathbf{2\times10^{-3}}$& $\mathbf{2\times10^{-7}}$ 	& $\mathbf{1.00}$  	& $\mathbf{10}$ & $\mathbf{0.036}$	& $\mathbf{1.0}$\\
      	Run B		& $1216^3$		& $6.6\times10^{-3}$	& $9\times10^{-6}$ 	& $0.12$ 			& $44$ 		& $1.1$			& $1.0$ \\
      	Run C		& $480^3$		& $1.5\times10^{-3}$	& $5\times10^{-6}$ 	& $0.12$ 			& $15$ 		& $0.27$		& $1.0$ \\
      	\textbf{Runs D05...10}	& $\mathbf{480^3}$		& $\mathbf{2\times10^{-3}}$	& $\mathbf{5\times10^{-6}}$ 	& $\mathbf{0.14}$ 			& $\mathbf{20}$ 		& $\mathbf{0.36}$		& $\mathbf{0.5...10}$ \\
      	Run E		& $480^3$		& $2.5\times10^{-3}$	& $5\times10^{-6}$ 	& $0.16$ 			& $25$ 		& $0.45$		& $1.0$ \\
      	Runs F1...10	& $576^3$		& $3\times10^{-3}$	& $5\times10^{-6}$ 	& $0.17$ 			& $30$ 		& $0.54$		& $0.5...10$ \\
      	Runs G05...10	& $480^3$		& $4\times10^{-3}$	& $2\times10^{-5}$ 	& $0.14$ 			& $20$ 		& $0.72$		& $0.5...10$ \\
      	Runs H05...10	& $480^3$		& $8\times10^{-3}$	& $2\times10^{-5}$ 	& $0.28$ 			& $20$ 		& $0.72$		& $0.5...10$ \\
      \hline
      \hline
    \end{tabular}
  \label{table}
\end{table}

\subsection{Reference run for chiral magnetically driven turbulence}

\begin{figure}
\begin{center}
  \includegraphics[width=\textwidth]{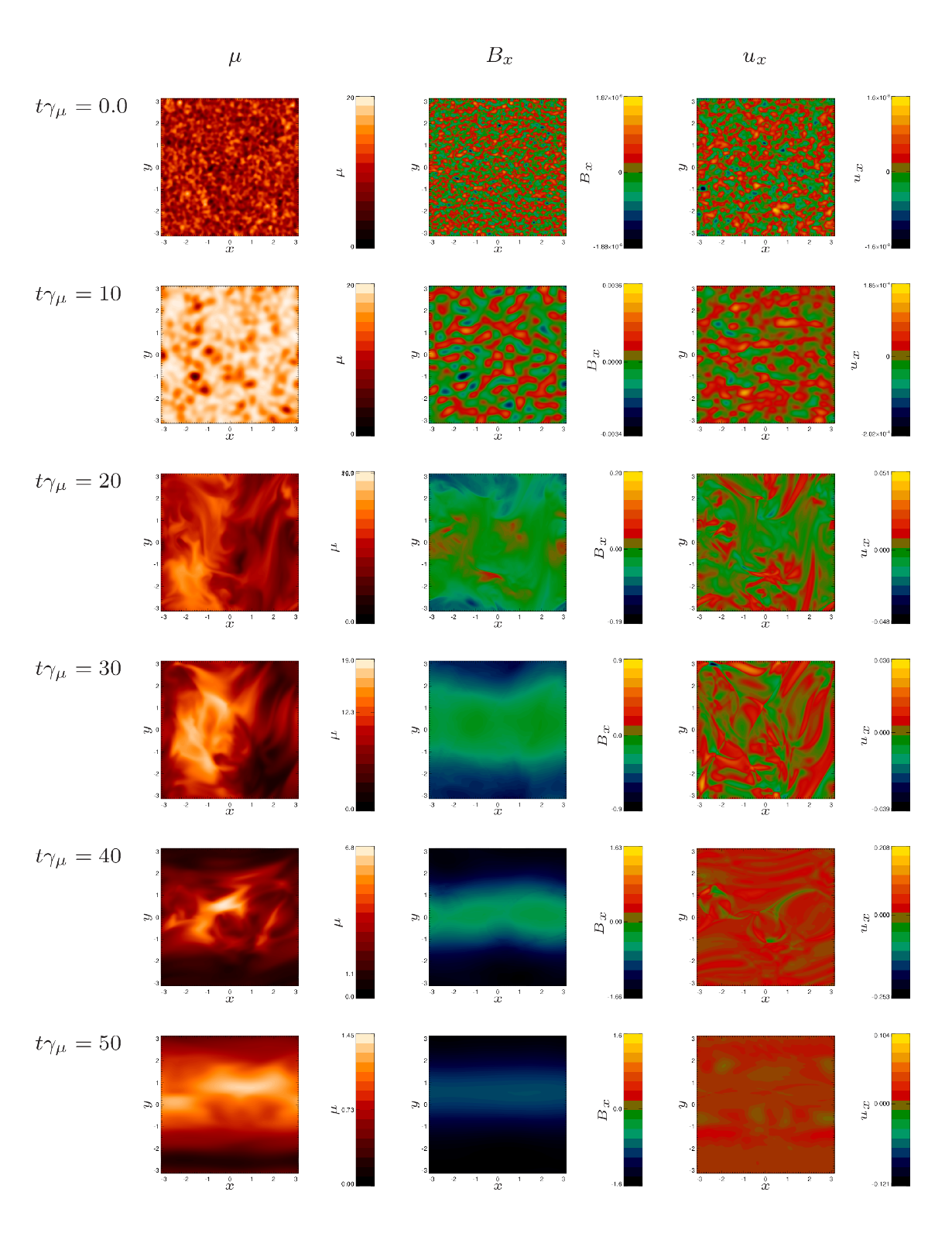}
\caption{Cross sections of the chiral chemical potential ($\mu$, left column),
as well as the $x$ components of the magnetic field ($B_x$, middle column),
and the velocity field ($u_x$, right column) in the $xy$ plane.
From top to bottom the time increases from $0$ to $50~\gamma_\mu^{-1}$,
during which the chiral MHD dynamo generates a large-scale magnetic field and
a velocity field.}%
\label{fig_snap}
\end{center}
\end{figure}

The generation of the magnetic field by the laminar chiral dynamo, and after that
by the chiral mean-field dynamo, can be seen in figure~\ref{fig_snap},
where we present snapshots of $\mu$ (left column), $B_x$
(middle column),
and  $u_x$ (right column) for run~A.
As indicated on the left, from top to bottom the time increases
from $t = 0$ to $t \gamma_\mu = 50$.
In the following, we summarise the quantitative analysis of run~A, but we refer
to section~4 of \citet{Schober2017} for a more detailed discussion of this
simulation.

\begin{figure}
\begin{center}
  \subfigure{\includegraphics[width=0.49\textwidth]{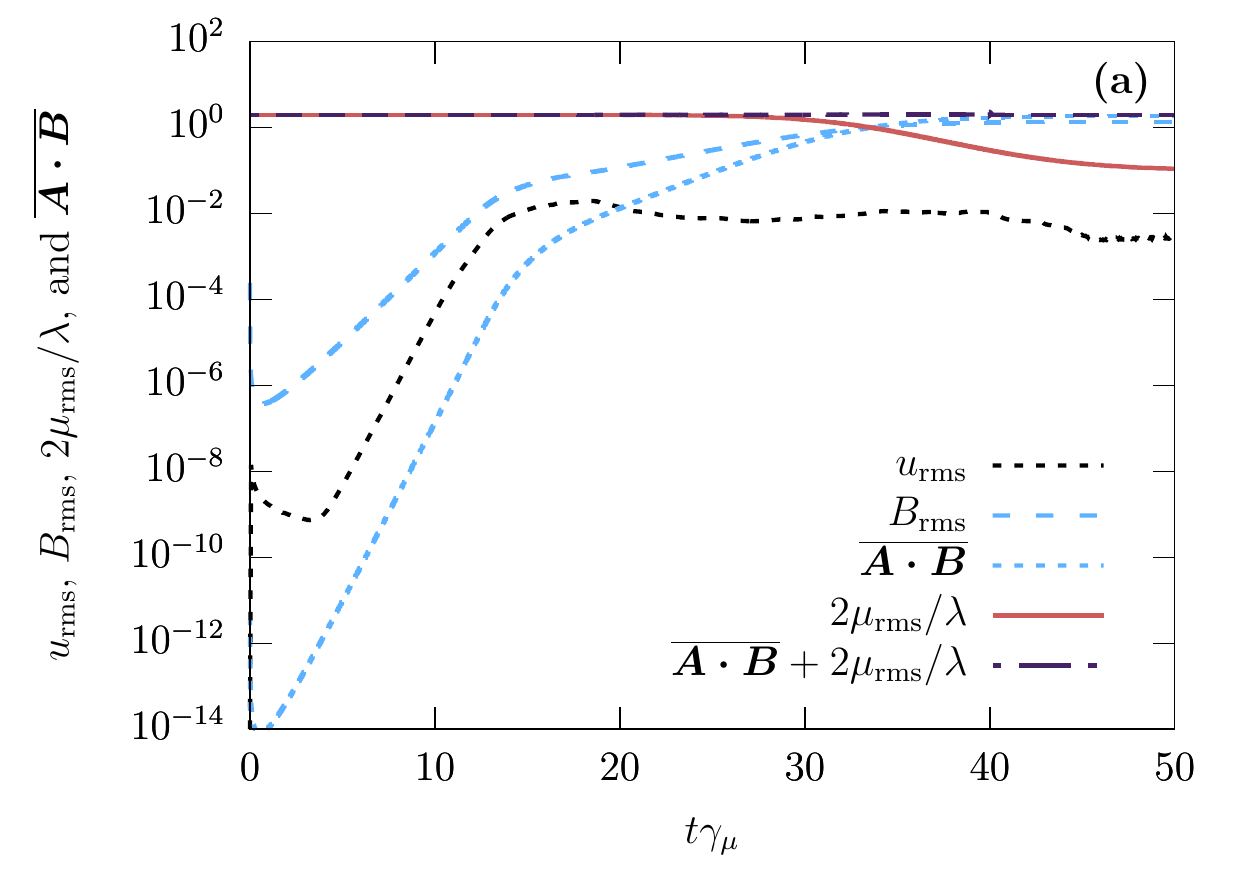}}
  \subfigure{\includegraphics[width=0.49\textwidth]{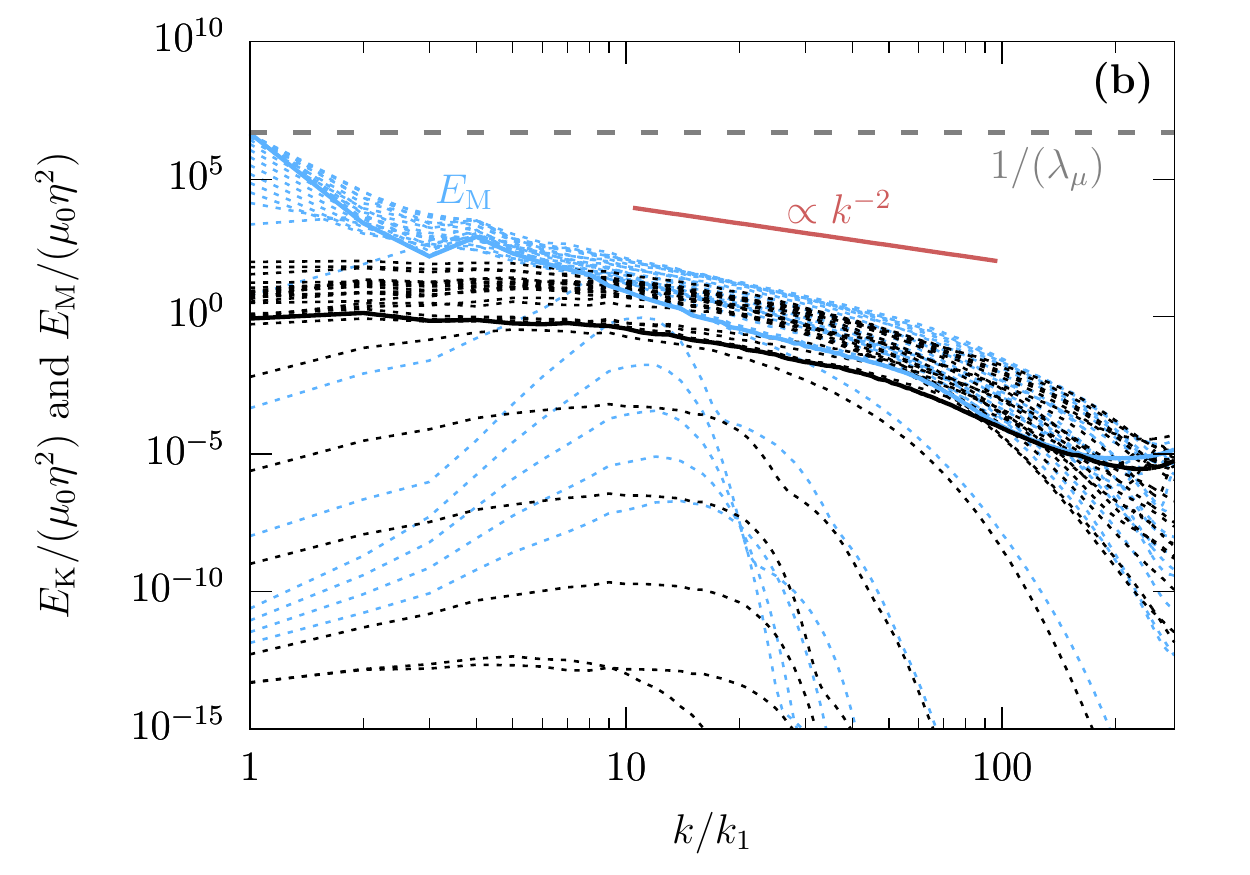}}
\caption{
(a)
The time evolution of the most relevant parameters in the reference run~A.
This run has been discussed in greater detail in section~4 of
\citet{Schober2017}, where this plot is shown in the top panel of their figure~9.
(b)
Normalized energy and helicity spectra for the reference run~A.
The time intervals between two spectra are equidistant and the last spectra are
presented by solid lines.
This plot is equivalent to figure~10 of \citet{Schober2017}. (colour online)
}
\label{fig_ts}
\end{center}
\end{figure}

The time evolution of the key quantities for the reference run~A
is shown in figure~\ref{fig_ts}.
As can be seen in figure~\ref{fig_ts}(a), $B_\mathrm{rms}$ (blue dashed line) increases
exponentially by over $4$ orders of magnitude before the growth rate decreases due
to the produced turbulence.
Saturation of the magnetic field growth occurs
at $t \gamma_\mu \approx 40$.
Both, the velocity $u_\mathrm{rms}$ (black dotted line) and the
magnetic helicity $\meanAB$ (blue dotted line) increase at a rate twice the one
of $B_\mathrm{rms}$.
The value of $\mu_\mathrm{rms}$ (orange solid line), here shown with a
constant factor $2/\lambda$,
decreases only above $t \gamma_\mu \approx 30$, switching off the chiral dynamo instability.
In accordance with the conservation law~(\ref{CL}), the sum
$\meanAB + 2\mu_\mathrm{rms}/\lambda$ (purple dashed-dotted line) is constant
throughout the simulation time.

In figure~\ref{fig_ts}(b), the evolution of the magnetic
(blue lines) and kinetic (black lines) energy spectra for run~A are presented.
It can be seen that the laminar chiral dynamo injects energy at the
wavenumber $k_\mu$, as given in equation~($\ref{eq_kmax}$).
Once turbulence has been produced,
the magnetic correlation length moves to smaller wavenumbers
due to mode coupling, similarly to what has been seen previously in
dynamos driven by the Bell instability \citep{RNBE2012}.
Eventually, the energy accumulates on $k=k_1$; see the final spectra in the
simulation which are plotted with solid lines.

\subsection{Turbulence in different scenarios}
\label{subsec_urmsBrms}

In chiral MHD, energy is transformed from the chiral chemical potential,
to magnetic energy, and later
to turbulent kinetic energy; see figure~\ref{fig_energyflow}.
For a quantification of this energy transfer,
it is useful to compare the production rate
of turbulent kinetic energy,
$\langle \UU \cdot (\JJ\times\BB)\rangle$, with the one of
the magnetic field, $|v_\mu \BB\cdot \nabla \times \BB|$.
Therefore, we define the dimensionless ratio
\begin{eqnarray}
   \Phi \equiv \frac{\langle \UU \cdot (\JJ\times\BB)\rangle}{|v_\mu \BB \cdot \nabla \times \BB|},
\label{eq_Phi}
\end{eqnarray}
where we have assumed that
$\BB\cdot \nabla \times \BB \approx k_\mathrm{M} B_\mathrm{rms}^2$ with the
inverse magnetic correlation length, $k_\mathrm{M}^{-1}$.

Furthermore, it is useful
to determine the value of turbulent kinetic energy
that can be produced in chiral MHD without external forcing of turbulence.
In the analysis of run~A, we have seen that the kinetic
energy reaches a certain percentage of the magnetic energy.
With the onset of the large-scale dynamo phase, phase 2, the ratio
\begin{eqnarray}
  \Upsilon \equiv \frac{\rho u_\mathrm{rms}^2/2}{B_\mathrm{rms}^2/2} 
\label{eq_Upsilon}
\end{eqnarray}
stays approximately constant and
decreases as soon as the peak of the magnetic energy spectrum reaches the
box wavenumber $k_1$.
Afterwards, turbulence is not driven by the Lorentz force anymore and
the velocity field decays.

In this section, we explore how the details of this scenario are affected by the
properties of the plasma.
In particular, we perform a parameter scan,
varying the chiral parameters as well as the magnetic Prandtl number.

\subsubsection{Dependence on the chiral parameters $\mathrm{Ma}_\mu$ and $\lambda_\mu$}

\begin{figure}
\begin{center}
  \subfigure{\includegraphics[width=0.49\textwidth]{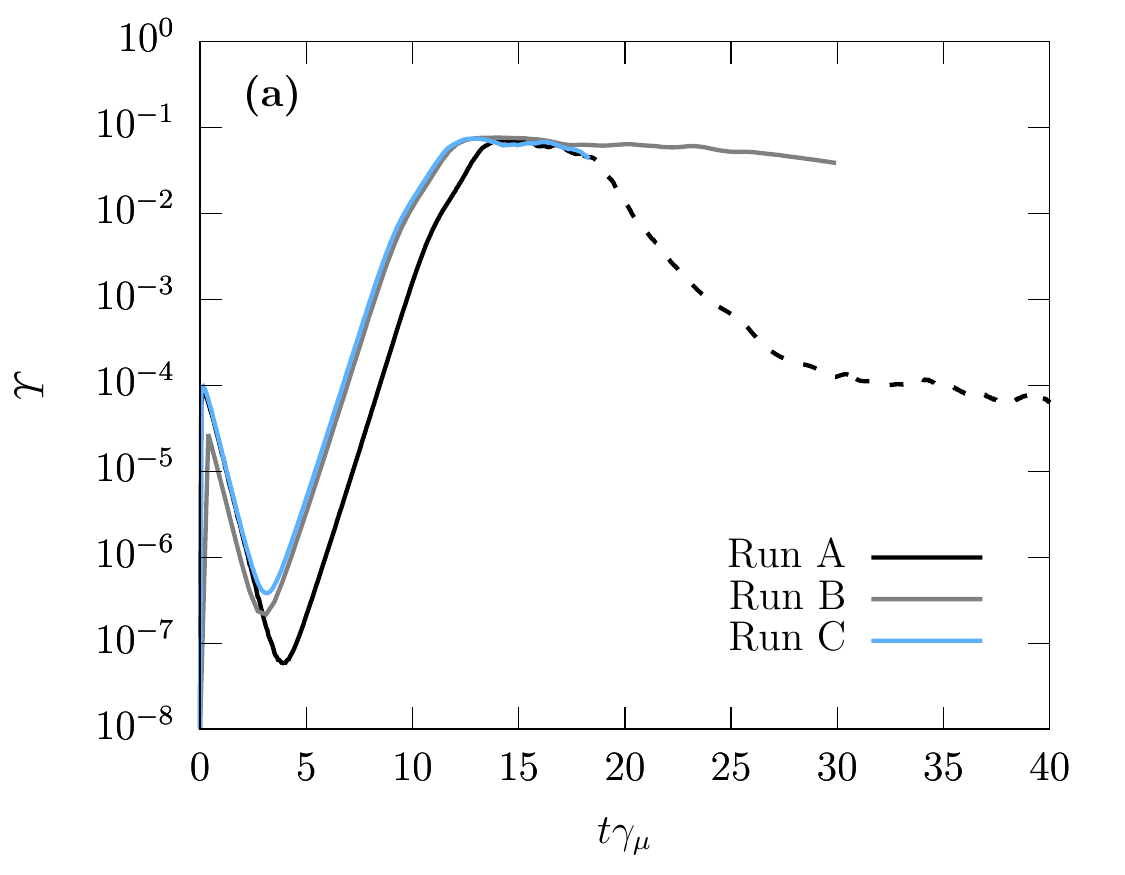}}
  \subfigure{\includegraphics[width=0.49\textwidth]{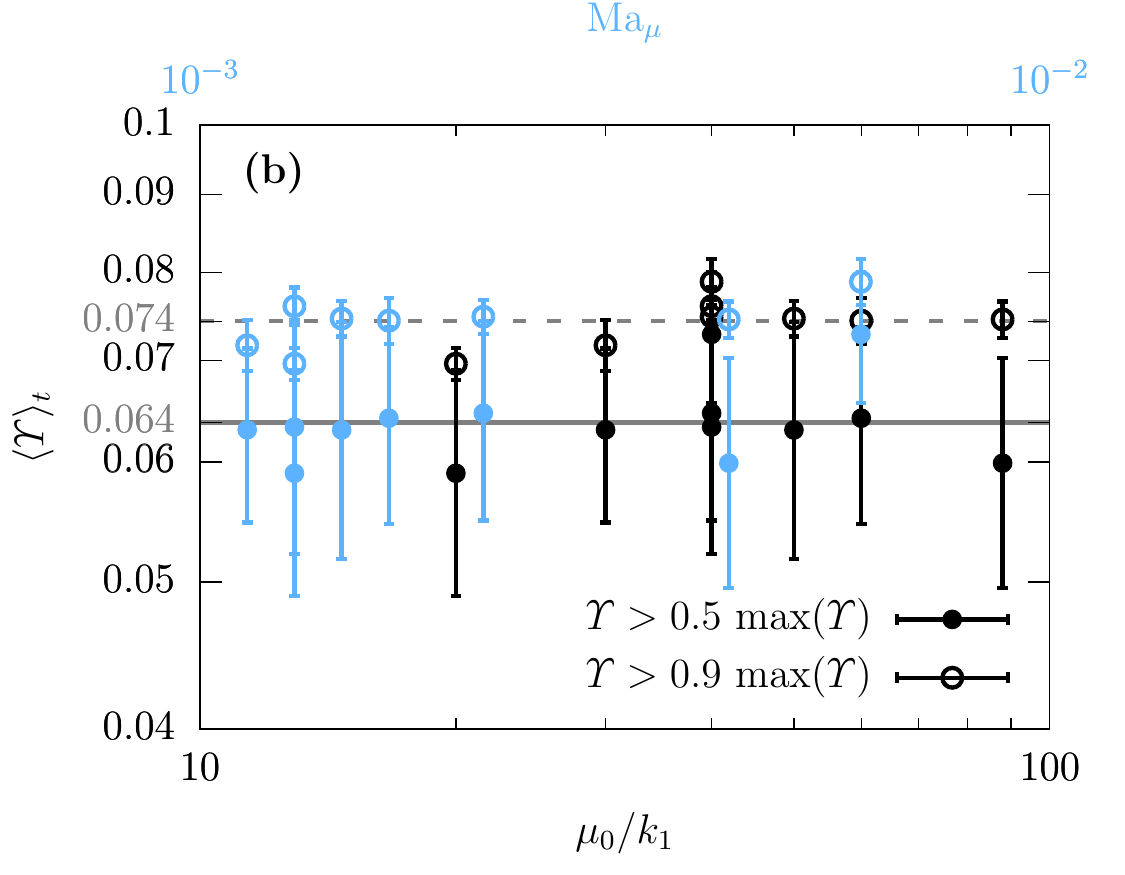}}
\caption{
(a)
The ratio of kinetic over magnetic
energy, $\Upsilon$, as a function of time,
normalized by $\gamma_\mu^{-1}$, for runs~A--C; see table~\ref{table}.
The time during which $k_\mathrm{M}$ is inside the numerical
box, i.e., $k_\mathrm{M}>k_1$, is marked by solid line style.
For $k_\mathrm{M}<k_1$, the lines are dashed.
(b)
The time averaged ratio
$\Phi$ as a function of $\mu_0/k_1$ (lower abscissa,
black symbols) and a function of $\mathrm{Ma}_\mu$ (upper abscissa, blue
symbols) for all runs with $\Pm=1$.
For the results shown as filled dots, the average has been performed for the
time interval for which $\Upsilon$ is at least 50\% of
its maximum value.
For open dots, the time average is taken for all
$\Upsilon> 0.9~\mathrm{max}(\Upsilon)$.
The solid grey line shows the mean value of $\UpsilonTimeAve$ resulting from the
first time averaging condition and the dashed grey for the latter condition.
(colour online)}%
\label{fig_urmsBrms_t}
\end{center}
\end{figure}

The time evolution of the ratio $\Upsilon$ is presented in 
figure~\ref{fig_urmsBrms_t}(a) for runs with different values of $\mathrm{Ma}_\mu$
and $\lambda_\mu$.
Time is normalised here by the inverse of the laminar dynamo growth
rate~(\ref{gamma-max}), allowing comparison between runs with different $v_\mu$.
The evolution of $\Upsilon$ in all runs is similar up to $t\approx 12\,\gamma_\mu^{-1}$,
except for a minor time delay of run~A.
This can be explained by the effect of magnetic diffusivity
which is larger than the one in run~C by a factor of two.
Phase 2, when turbulence affects the evolution of the magnetic field, begins
approximately at
$t\approx (12$--$14)~\gamma_\mu^{-1}$
for the runs considered here.
The onset of phase 2 is weakly dependent on $\eta$ and, in principle, also on
the initial value of the magnetic field strength, which is the same for all
runs presented in this paper.
During phase 2, the ratio $\Upsilon$ is comparable for all
three runs considered here, even
though $\mathrm{Ma}_\mu$ and $\lambda_\mu$ are different.
Once the chiral large-scale dynamo phase begins,
we obtain the ratio $\Upsilon \lesssim 0.1$.

Run~A, the reference run discussed in the previous section,
has the lowest value of $\lambda$ in our sample, leading to a small value of
$k_\lambda$ in comparison to the maximum wavenumber in the box:
$k_\lambda \approx 0.036 k_1$.
This implies that $k_1$ is reached early, much before
dynamo saturation, and the kinetic energy decays.
As long as $k_\mathrm{M} > k_1$, a solid line style is used in
figure~\ref{fig_urmsBrms_t}, while for late times,
the time evolution is presented with dashed lines
to indicate the finite box effect.
Here, $k_\mathrm{M}$ has been determined as the peak of the energy
spectra.
To observe a scenario in which the complete inverse cascade takes place inside the box,
so that the kinetic energy does not decay within phase 2, we have performed run~B.
The choice of the parameters for this run, listed in table~\ref{table}, can be
justified as follows:
As for all of the runs presented in this paper, the ratio between $k_\mu$ and
$k_\lambda$ needs to be large, in order to observe
a chiral large-scale dynamo phase (phase 2).
Using equations~(\ref{eq_kmax}) and (\ref{klambda}), we see that
\begin{equation}
   \frac{k_\mu}{k_\lambda} =  \left(\frac{C_\lambda}{4 C_\mu} \frac{1}{\meanrho \eta^2 \lambda}\right)^{1/2}
                           = \left(\frac{C_\lambda}{4 C_\mu} \frac{1}{\lambda_\mu}\right)^{1/2},
\label{eq_kmuklambda}
\end{equation}
which is independent of $\mu_0$.
However, $\mu_0$ needs to be chosen high enough to ensure that $k_\lambda>k_1$.
In run~B, we use $\mu_0/k_1 = 88$, which implies that the laminar dynamo
instability occurs on small spatial scales and a high numerical resolution
is required.
Run~B is presented in figure~\ref{fig_urmsBrms_t}(a) as a grey
solid line, for which the ratio $\Upsilon$ remains
approximately constant for times larger than $\approx 12~\gamma_\mu^{-1}$.

In figure~\ref{fig_urmsBrms_t}(b), we show the time averaged ratio
$\UpsilonTimeAve$ for all our runs with $\Pm=1$ as a
function of $\mu_0/k_1$ as black symbols.
The blue symbols refer to the upper $x$ axes and indicate the
corresponding value of $\mathrm{Ma}_\mu$.
For the time averaging procedure, we consider two different criteria: For solid
symbols the time average is performed for all values of
$\Upsilon$ larger than 50\% of its maximum value.
The open dots are obtained by using all values for which
$\Upsilon > 0.9~\mathrm{max}(\Upsilon)$,
which obviously results in a larger average value.
Error bars represent the standard deviation of $\UpsilonTimeAve$.
There is no significant dependence of $\UpsilonTimeAve$
on the values of $\mu_0$ and $\mathrm{Ma}_\mu$ for the parameter space explored
here:
When averaging over all $\Upsilon > 0.5~\mathrm{max}(\Upsilon)$, we find a mean
$\UpsilonTimeAve\approx0.064$, and when employing the criterion
$\Upsilon > 0.9~\mathrm{max}(\Upsilon)$, we find $\UpsilonTimeAve\approx0.074$.

\begin{figure}
\begin{center}
  \subfigure{\includegraphics[width=0.49\textwidth]{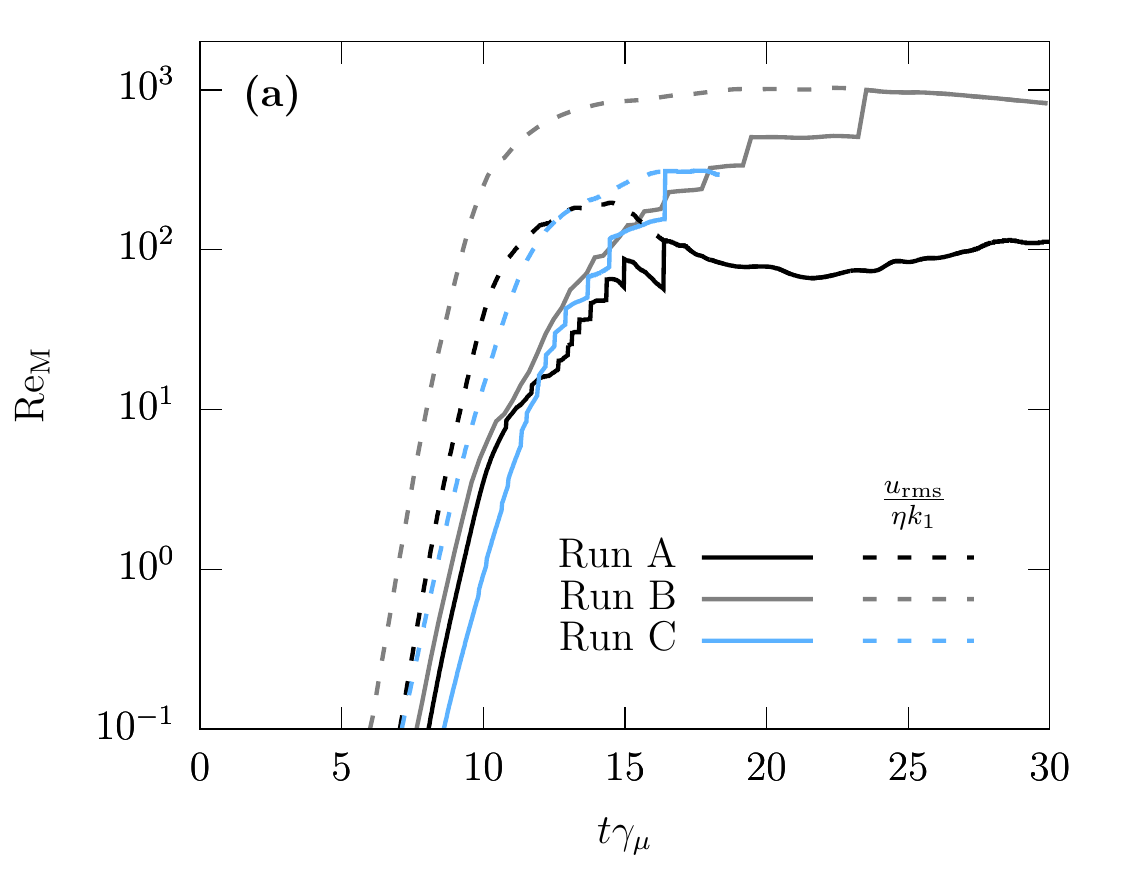}}
  \subfigure{\includegraphics[width=0.49\textwidth]{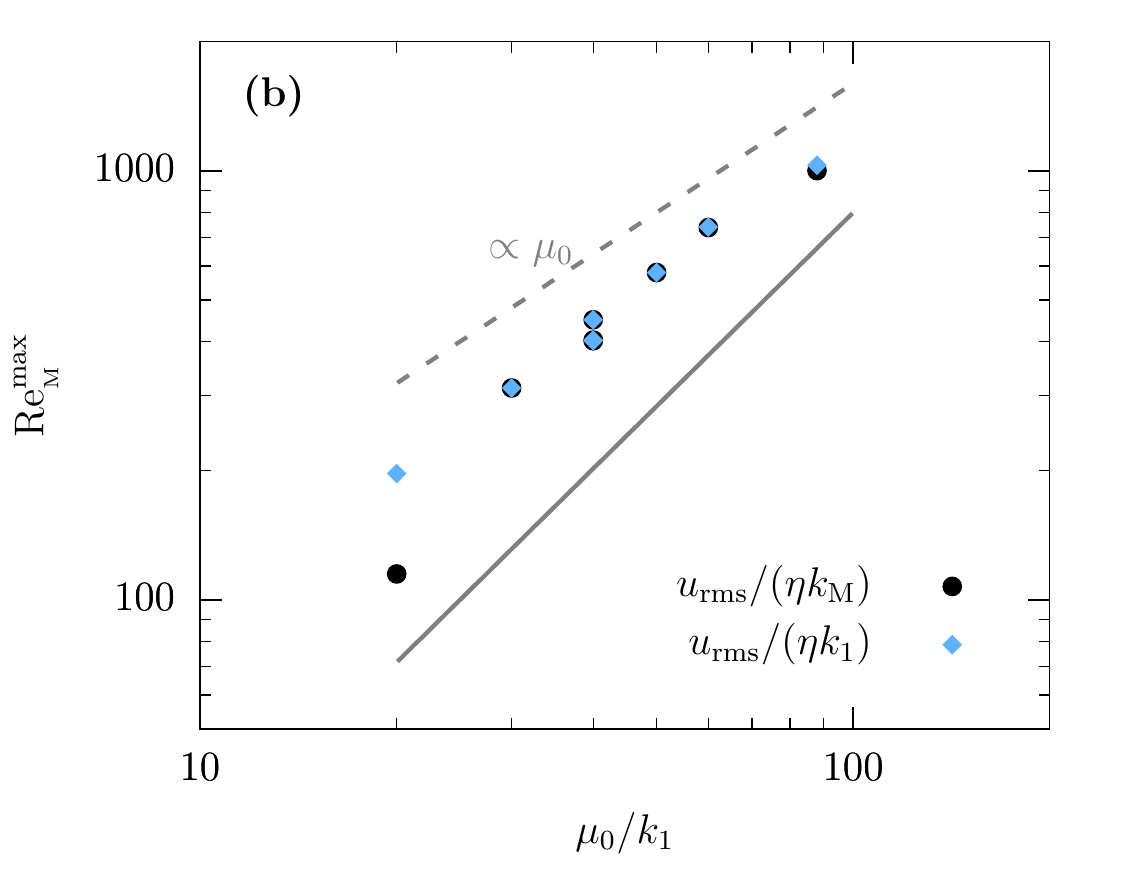}}
\caption{
(a)
The magnetic Reynolds number $\Rm$ as a function of time for runs~A--C.
The solid lines indicate the result using $k_\mathrm{M}$ as the integral scale
of turbulence and dashed lines show the result using $k_1$.
At late times $k_\mathrm{M}=k_1$ in our DNS, which are confined within a finite
box.
(b)
The maximum magnetic Reynolds number found in our
simulations versus $\mu_0/k_1$, for both cases $\Rm = u_\mathrm{rms}/(\eta k_1)$
(black dots) and $\Rm = u_\mathrm{rms}/(\eta k_\mathrm{M})$ (blue diamonds).
The solid line indicates the scaling $\mu_0^{3/2}$ which is expected for finite
box simulations with $k_1>k_\lambda$. (colour online)
}
\label{fig_Rmmax_mu}
\end{center}
\end{figure}

To estimate the magnetic Reynolds number, we need to determine
the amount of turbulent kinetic energy that can be produced by the
Lorentz force.
The value of $\Rm$ is determined by the rms velocity, the magnetic diffusivity, and
the correlation length of the magnetic field, $k_\mathrm{M}^{-1}$.
In numerical simulations, both $u_\mathrm{rms}$ and $k_\mathrm{M}$ can be
limited by the size of the box, while $\eta$ is an input parameter.
We have seen that $\UpsilonTimeAve$ has an
approximately fixed value in the mean-field dynamo phase for $\Pm=1$ and as
long as $k_\mathrm{M} > k_1$.
This implies that the value of $u_\mathrm{rms}$ is proportional to
$B_\mathrm{rms}$ and reaches a maximum at the time $t_\mathrm{box}$,
which is defined as the time
when the peak of the energy spectrum reaches the size of the box,
i.e., when $k_\mathrm{M}= k_1$.
The energy spectrum at this time,
described by equation~(\ref{Cmu}), reaches a maximum
$E_\mathrm{M}(k_1, t_\mathrm{box}) = C_\mu \meanrho \eta^2 \mu_0^3/2$.
The magnetic field strength corresponding to this energy spectrum can be
estimated as $[E_\mathrm{M}(k_1, t_\mathrm{box})k_1]^{1/2}
= (C_\mu \meanrho/2)^{1/2} \eta \mu_0^{3/2}$.
Hence we expect a scaling of the maximum velocity in our simulations
$\propto \eta \mu_0^{3/2}$.
The magnetic Reynolds number, $\Rm = u_\mathrm{rms}/(\eta k_\mathrm{M})$ with
$k_\mathrm{M}=k_1$ at late times, is thus independent of $\eta$
and scales as $\mu_0^{3/2}$.
This scaling is observed in our simulations; see figure~\ref{fig_Rmmax_mu}(b).
The Reynolds number as a function of time for runs with different $\mu_0/k_1$
is presented in figure~\ref{fig_Rmmax_mu}(a).
Here, we show two different ratios, $u_\mathrm{rms}/(\eta k_\mathrm{M})$ and
$u_\mathrm{rms}/(\eta k_1)$.
For the first case, $k_\mathrm{M}$ is measured as a function of time using the
magnetic energy spectra.
At late times, once the peak of the magnetic energy spectrum has reached the
box wavenumber, $k_\mathrm{M}=k_1$, and the dashed and solid curves for individual
runs coincide.

\begin{figure}
\begin{center}
  \subfigure{\includegraphics[width=0.49\textwidth]{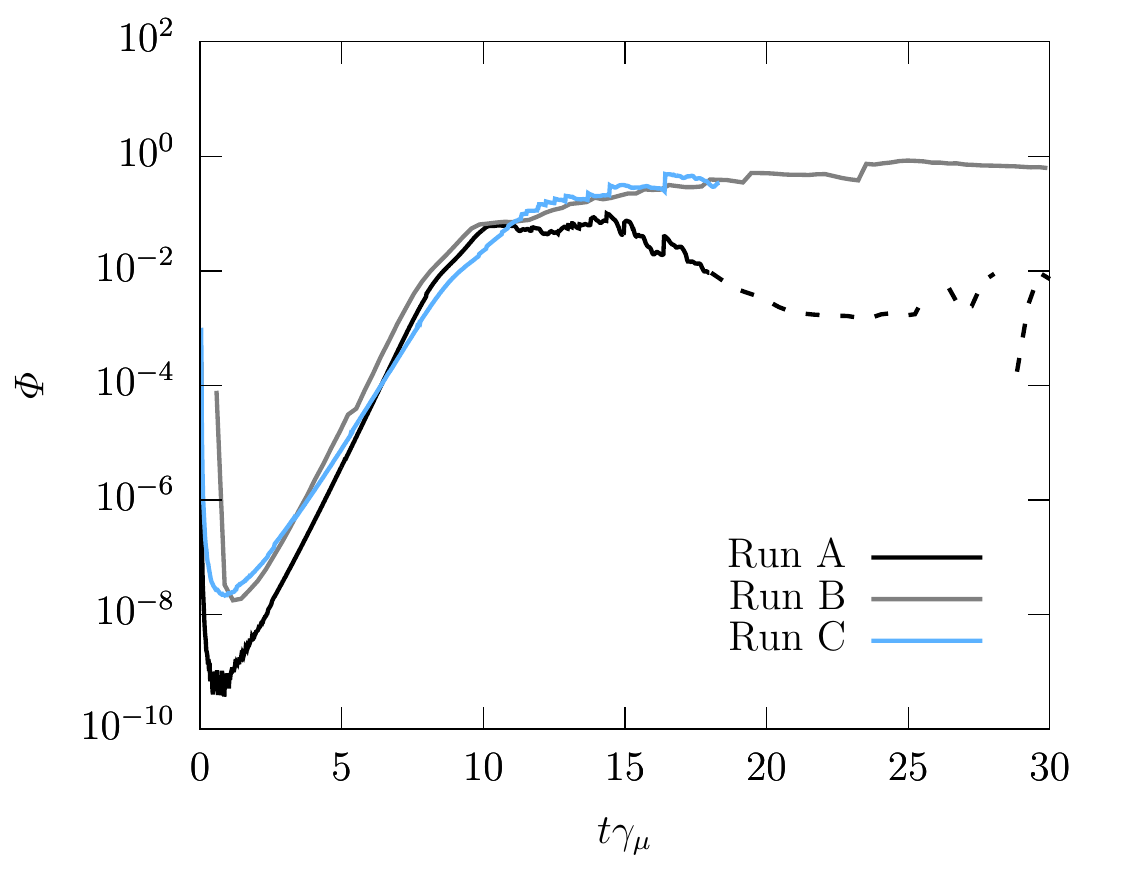}}
\caption{
Same as figure~\ref{fig_urmsBrms_t}(a), but for the ratio of
the production rate of kinetic energy over the production rate of magnetic 
energy, $\Phi$. (colour online)
}
\label{fig_Phi_t}
\end{center}
\end{figure}

The ratio of the production rate of kinetic energy over the production
rate of magnetic energy, $\Phi$, is presented in
figure~\ref{fig_Phi_t}.
For runs~A, B, C, it can be seen how $\Phi$ increases exponentially with a
rate $\approx 2\gamma_\mu$ like the velocity field.
In phase~2, the growth rate decreases, and $\Phi$ seems to converge to a value
of $\approx 1$ at
dynamo saturation, implying that the transfer rate from the chiral
chemical potential to magnetic energy and the transfer rate from magnetic
energy to kinetic energy become comparable.
Again, run~A differs from runs~B and C, due to the fact that the magnetic
correlation length reaches the size of the domain at $t \gamma_\mu \approx 17$,
which is well before dynamo saturation.
In order to perform a quantitative analysis of the dependence of $\Phi$ on
the plasma parameters, it would be necessary to run the numerical simulations
for a much longer time,
i.e., until dynamo saturation, and to increase the size of
the simulation domain to ensure $k_1<k_\mathrm{M}$.
This is, however, too expensive and beyond the scope of this paper.

\subsubsection{Dependence on the magnetic Prandtl number}
\label{sec_PmDependence}

\begin{figure}
\begin{center}
  \subfigure{\includegraphics[width=0.49\textwidth]{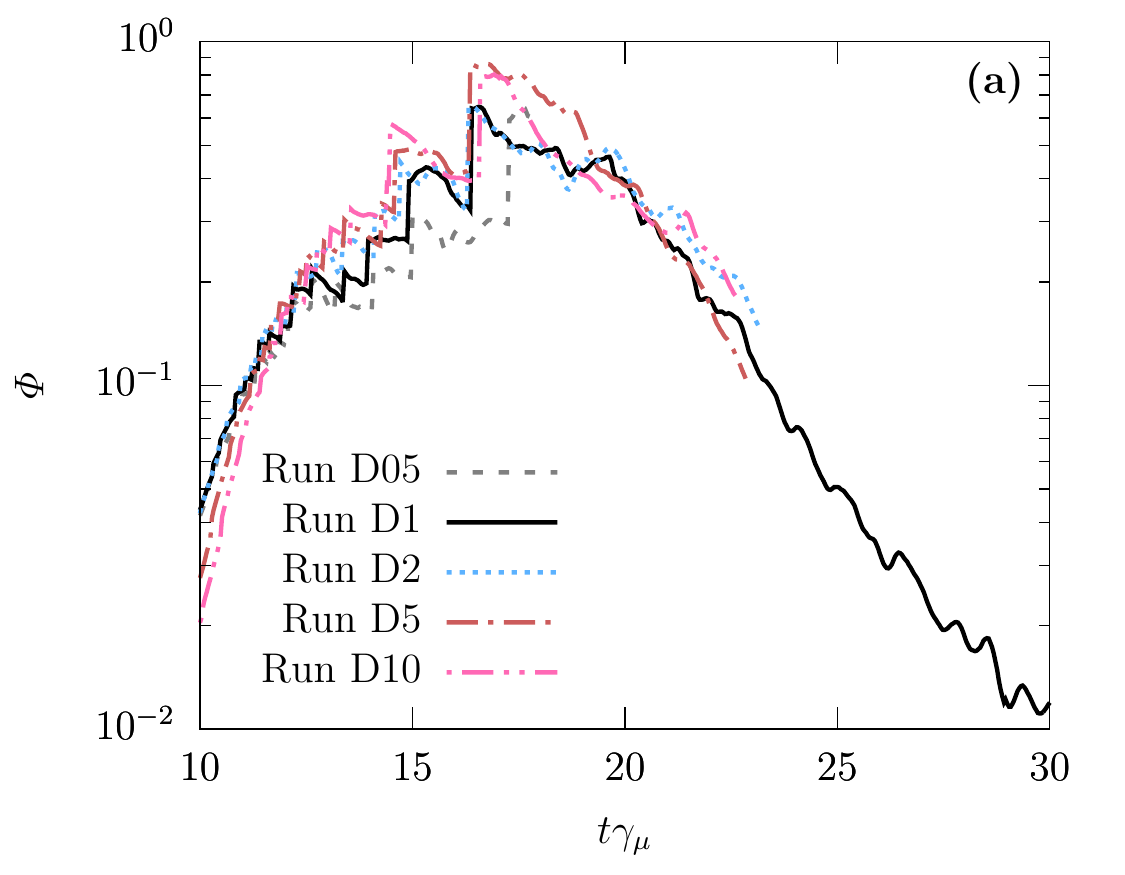}}
  \subfigure{\includegraphics[width=0.49\textwidth]{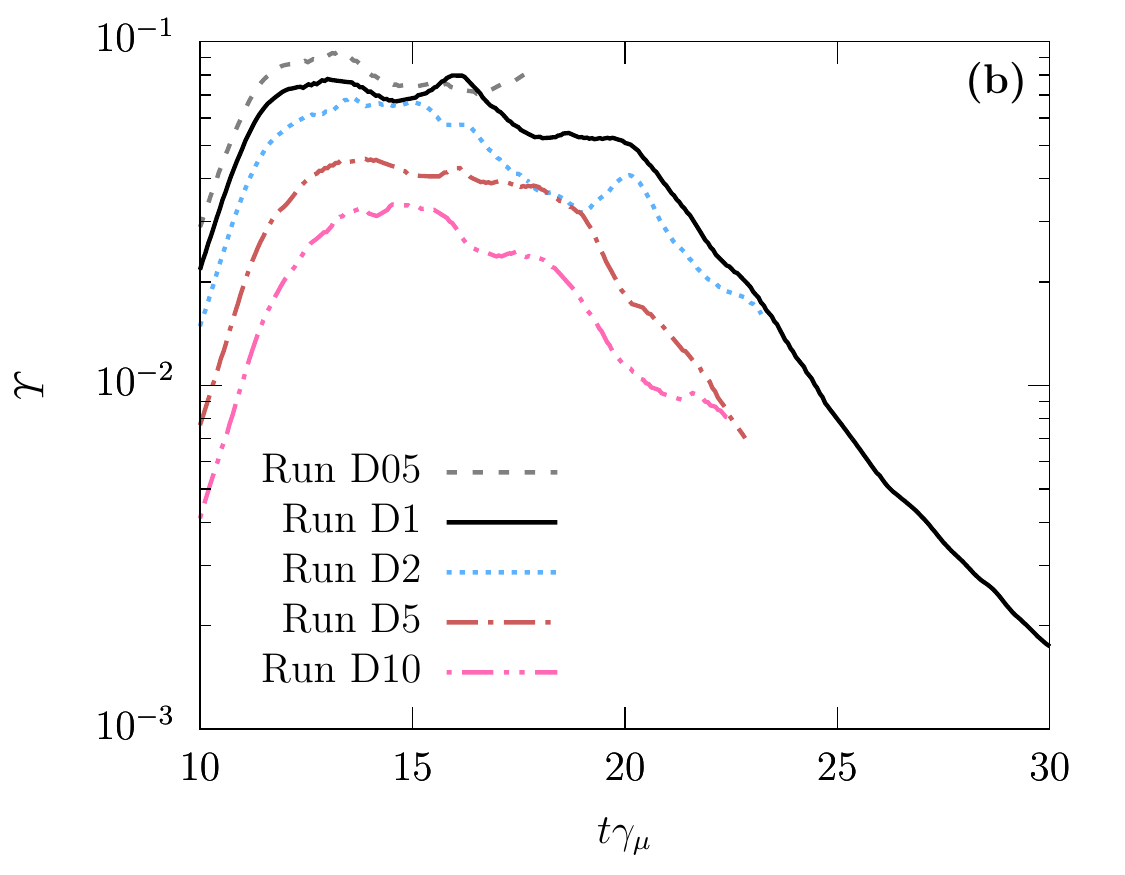}}
  \subfigure{\includegraphics[width=0.49\textwidth]{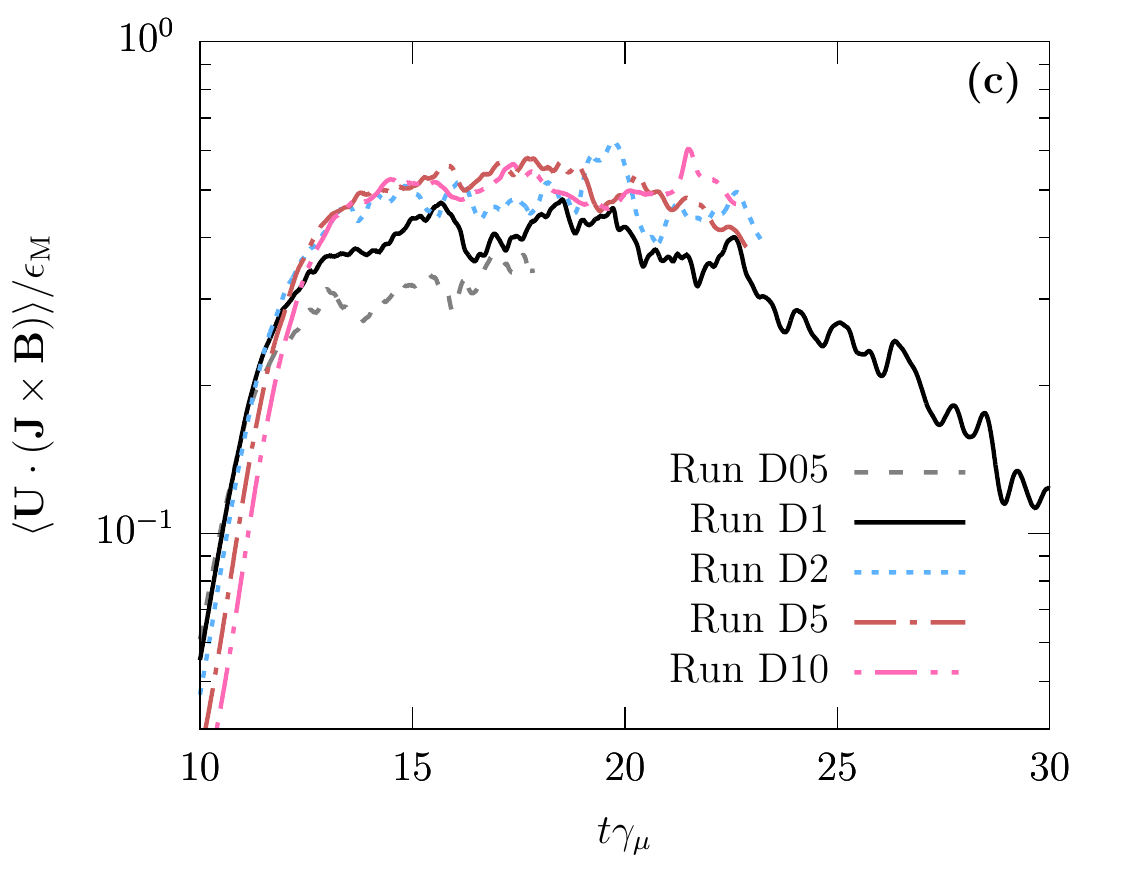}}
  \subfigure{\includegraphics[width=0.49\textwidth]{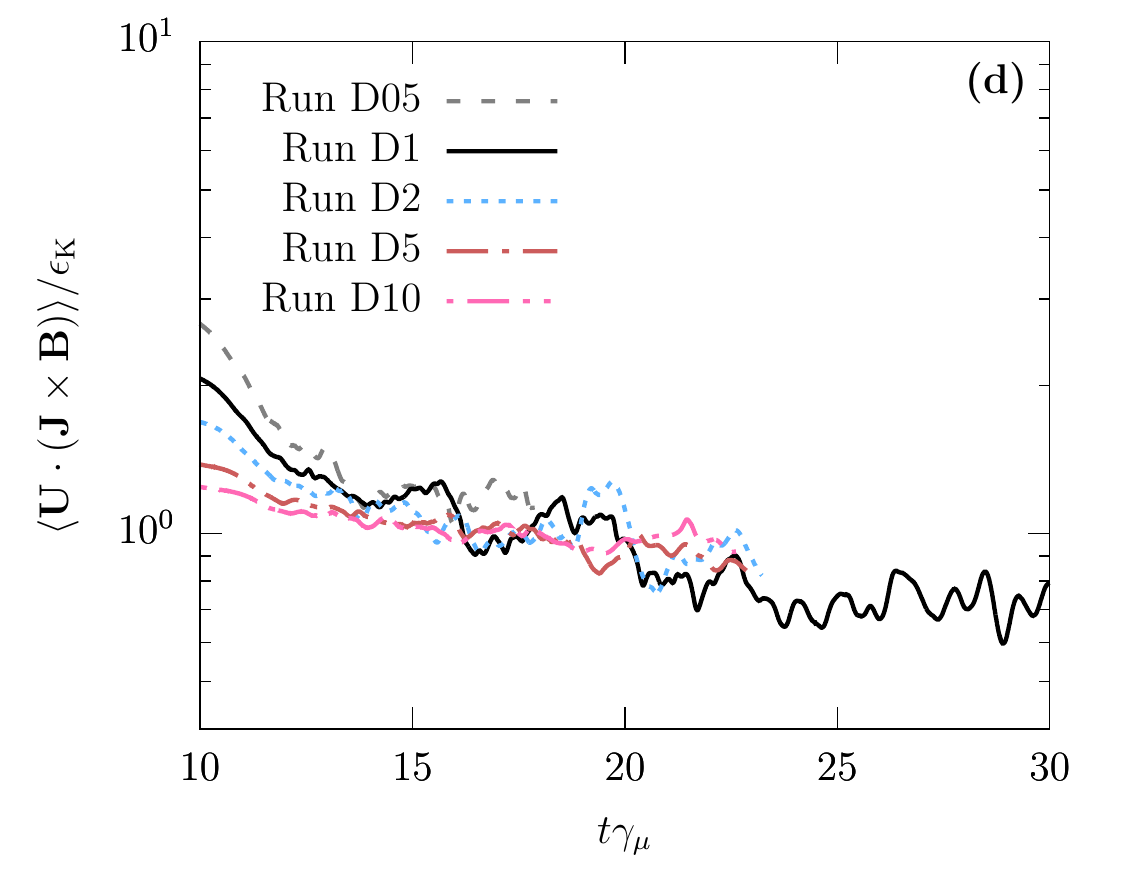}}
  \subfigure{\includegraphics[width=0.49\textwidth]{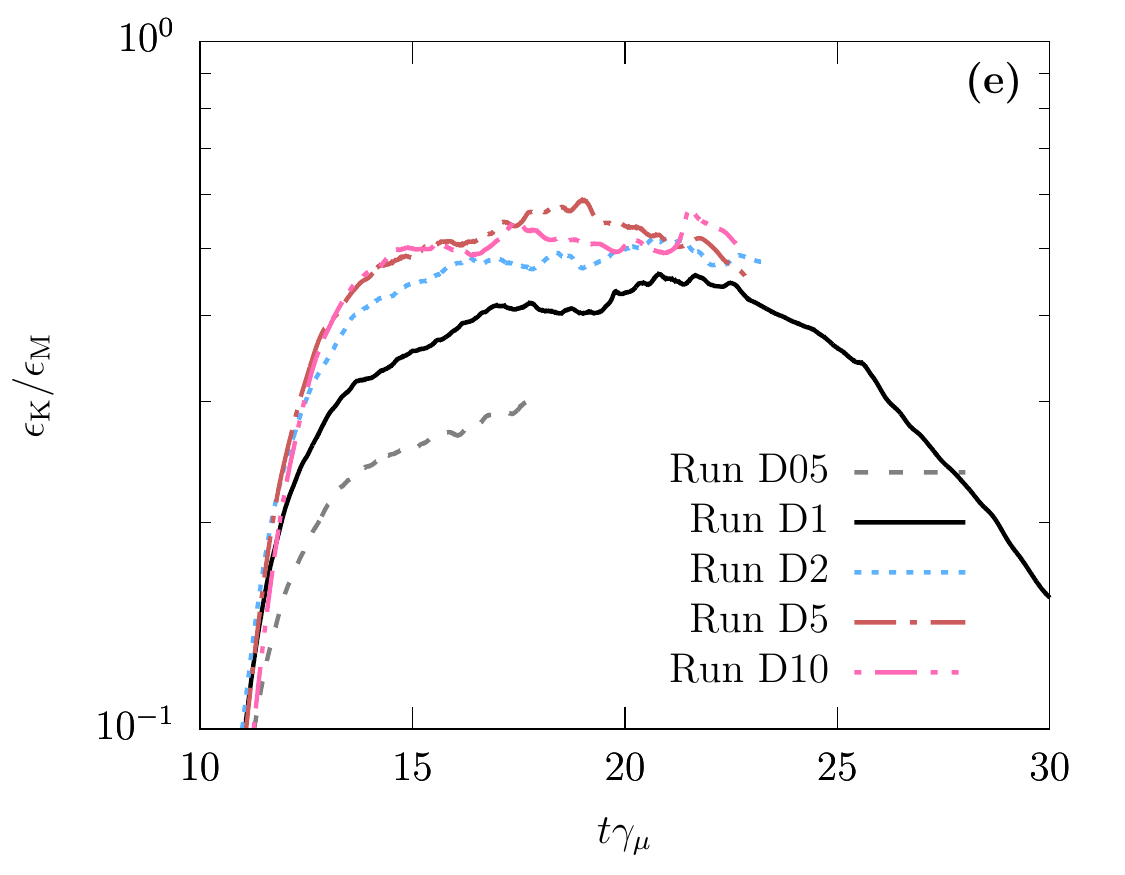}}
  \subfigure{\includegraphics[width=0.49\textwidth]{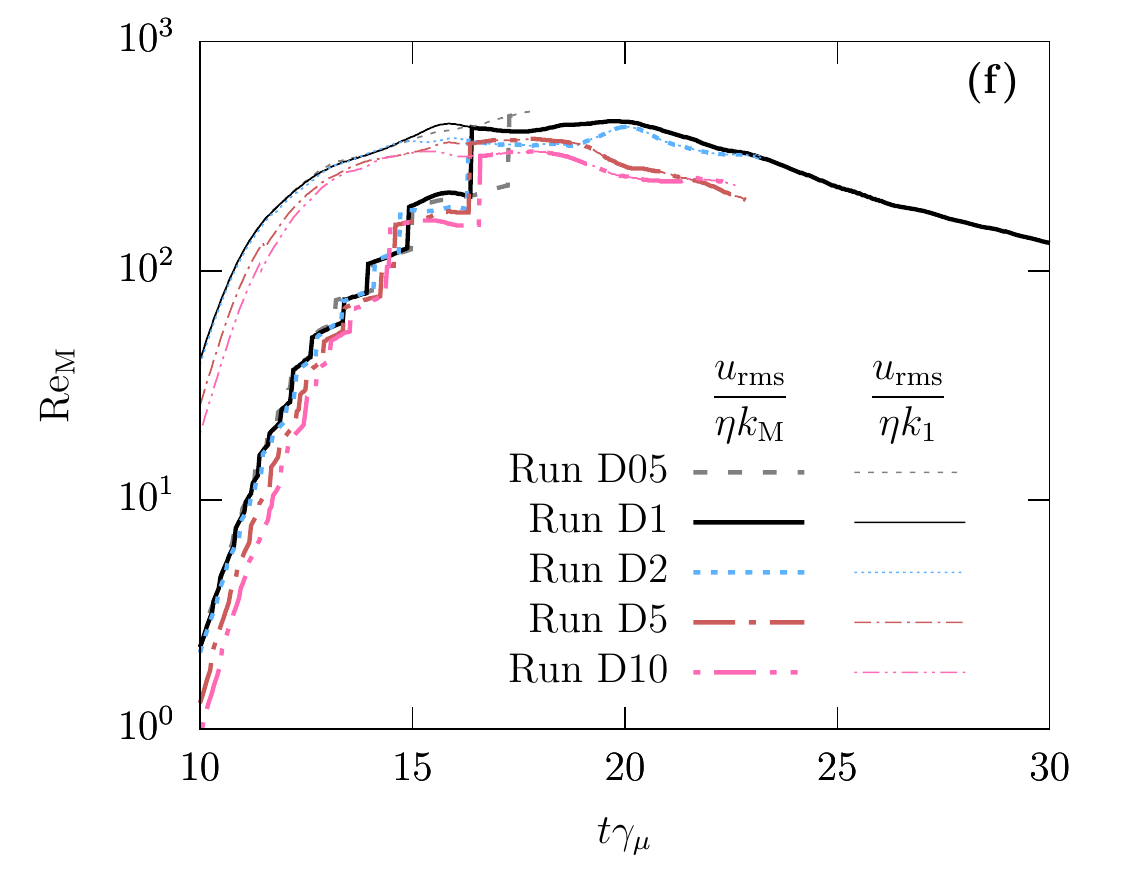}}
\caption{
Time series of the ratio of 
(a) energy transfer rates $\Phi$,
(b) the energy ratio $\Upsilon$,
(c) $\langle \UU \cdot (\JJ\times\BB)\rangle/\epsilon_\mathrm{M}$,
(d) $\langle \UU \cdot (\JJ\times\BB)\rangle/\epsilon_\mathrm{K}$,
(e) $\epsilon_\mathrm{K}/\epsilon_\mathrm{M}$, and
(f) $\Rm$ for series~D; see table~\ref{table}.
(colour online)}
\label{fig_urmsBrms_t__Pm}
\end{center}
\end{figure}

We have performed a series of runs with different magnetic Prandtl numbers by
changing the value of $\nu$, in order to explore
trends in the conversion of magnetic to kinetic energy.
The time series of the most relevant quantities are discussed here
exemplary for the runs of series~D, where $\Pm$ varies between $0.5$ and $10$ while
all other run parameters are unchanged; see table~\ref{table}.
It should be noted that low $\Pm$ are difficult to obtain in DNS at fixed
$\eta$, as an increase of resolution is required when decreasing $\nu$,
making a quantitative study of the low Prandtl number regime inaccessible
to our current simulations.

\begin{figure}
\begin{center}
  \includegraphics[width=0.49\textwidth]{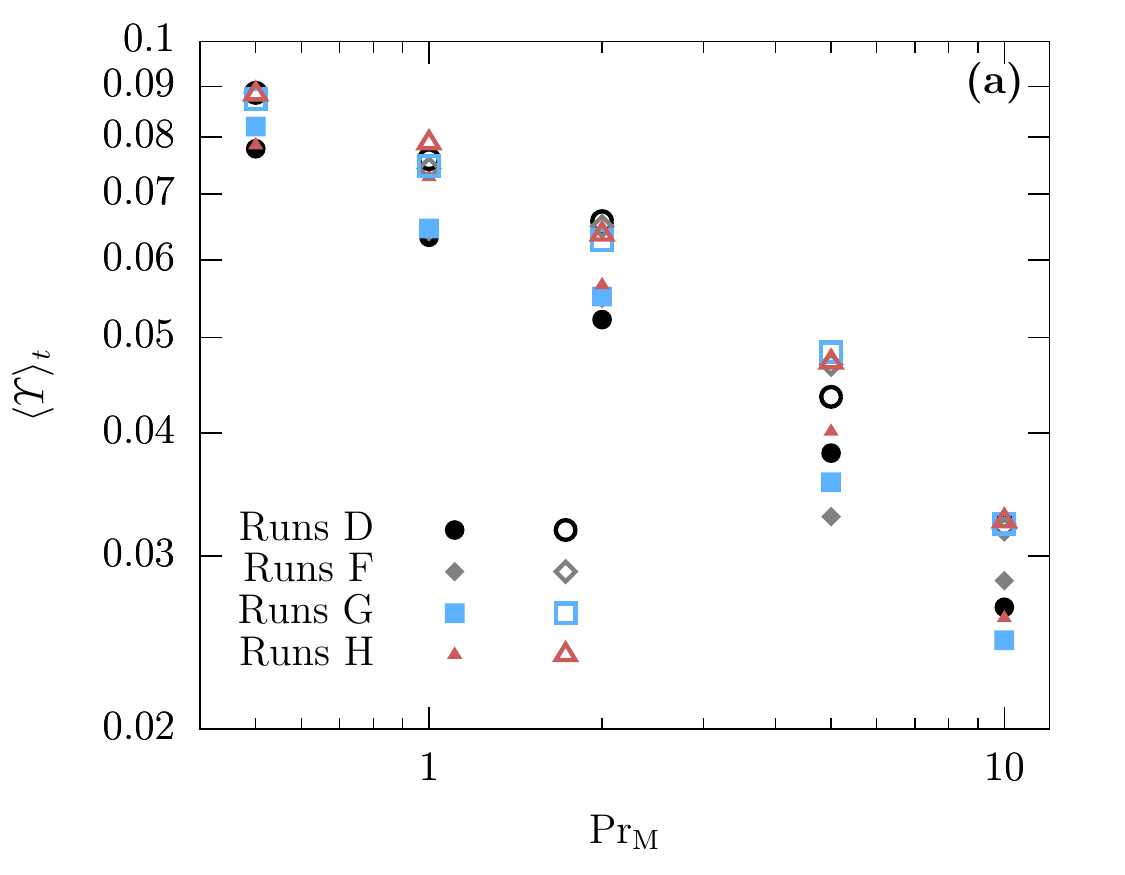}
  \includegraphics[width=0.49\textwidth]{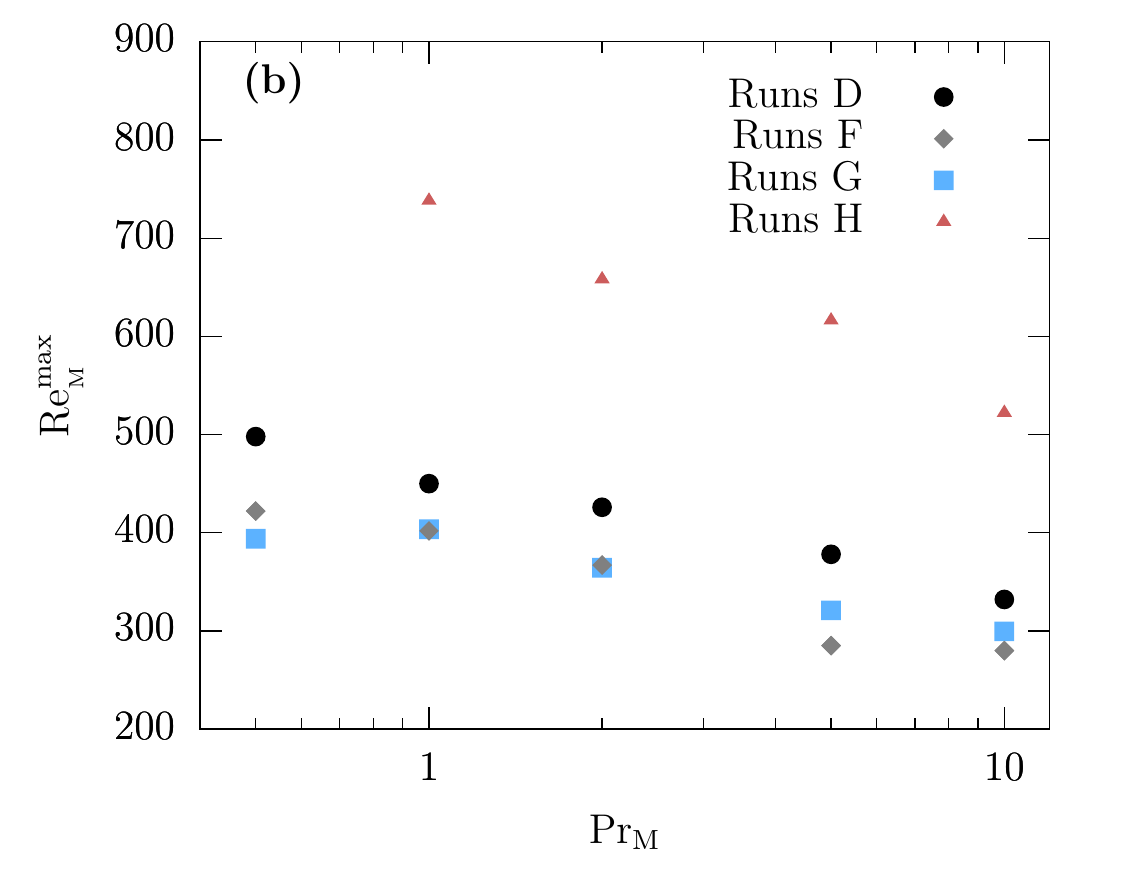}
  \includegraphics[width=0.49\textwidth]{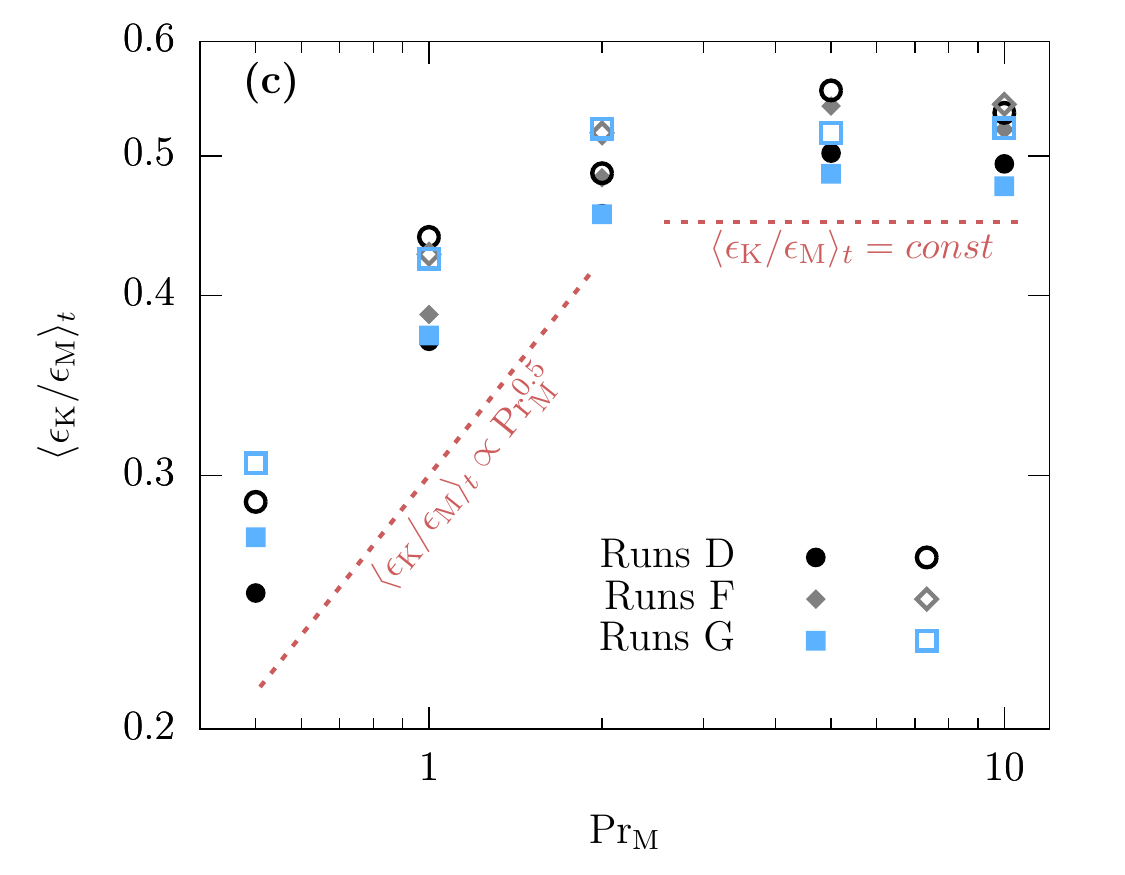}
\caption{
(a)
The time averaged value of $\Upsilon$
for different run series.
Filled symbols represent the result from an time averaging of all data points
with $\Upsilon > 0.5~\mathrm{max}(\Upsilon)$
and open symbols averaging is performed for all data with
$\Upsilon > 0.9~\mathrm{max}(\Upsilon)$.
Errors are of the the order of 10\% for the filled symbols and 5\% for the open
symbols, but not presented in the figure for better visualization.
(b)
The maximum magnetic Reynolds number in the different DNS as a function of $\Pm$. 
(c)
The time averaged ratio of viscous over Joule dissipation,
$\langle\epsilon_\mathrm{K}/\epsilon_\mathrm{M}\rangle_t$, as a function of $\Pm$.
Again, for filled symbols, the average of
$\epsilon_\mathrm{K}/\epsilon_\mathrm{M}$ is calculated for all values
$\epsilon_\mathrm{K}/\epsilon_\mathrm{M} > 0.5~\mathrm{max}(\epsilon_\mathrm{K}/\epsilon_\mathrm{M})$
and for open symbols, averaging is performed for
$\epsilon_\mathrm{K}/\epsilon_\mathrm{M} > 0.9~\mathrm{max}(\epsilon_\mathrm{K}/\epsilon_\mathrm{M})$.
(colour online)}
\label{fig_urmsBrms_Pm__mean}
\end{center}
\end{figure}

In figure~\ref{fig_urmsBrms_t__Pm}(a), $\Phi$ is presented
as a function of time.
It should be noted, that the magnetic correlation length reaches $k_1^{-1}$
at a time of $t\gamma_\mu \approx 20$.
Later times should not be discussed due to numerical artefacts discussed before.
Up to $t\gamma_\mu \approx 20$, we do not observe a significant $\Pm$-dependence
of $\Phi$.
The ratio $\Upsilon$, presented 
in figure~\ref{fig_urmsBrms_Pm__mean}(b),
on the other hand decreases significantly with increasing $\Pm$.

In the middle panels (c),(d) of figure~\ref{fig_urmsBrms_t__Pm}, we present the time
evolution of
additional characteristics describing the transfer from
kinetic to magnetic energy.
We find that both, the work done by the Lorentz force,
$\langle\mathbf{U}\cdot(\mathbf{J}\times\mathbf{B})\rangle$, and the
ratio of viscous over Joule dissipation,
$\epsilon_\mathrm{K}/\epsilon_\mathrm{M}$ depend on $\Pm$.
The latter dissipation ratio is expected to increase with $\Pm$ for large-scale
and small-scale dynamos in classical MHD; see \citet{B14}.
This trend of $\epsilon_\mathrm{K}/\epsilon_\mathrm{M}$
with $\Pm$ is also observed in our DNS
of chiral large-scale dynamos; see also 
figure~\ref{fig_urmsBrms_Pm__mean}(c).
However, we do observe a power-law scaling of
$\langle\epsilon_\mathrm{K}/\epsilon_\mathrm{M}\rangle_t$ with $\Pm$ only
below $\Pm\approx 2$.
For larger Prandtl numbers the ratio becomes constant.

The maximum magnetic Reynolds numbers, $\Rm^\mathrm{max}$, obtained in series~D,
are presented figure~\ref{fig_urmsBrms_Pm__mean}(b).
It can be seen that a dependence of $\Rm^\mathrm{max}$ on $\Pm$
is caused by the decrease of $u_\mathrm{rms}$ with increasing $\Pm$.
As discussed in the previous section, a change of $\Rm$ in DNS with
$k_\lambda< k_1$ can only be achieved by changing $\mu_0$.

\begin{figure}
\begin{center}
  \includegraphics[width=0.49\textwidth]{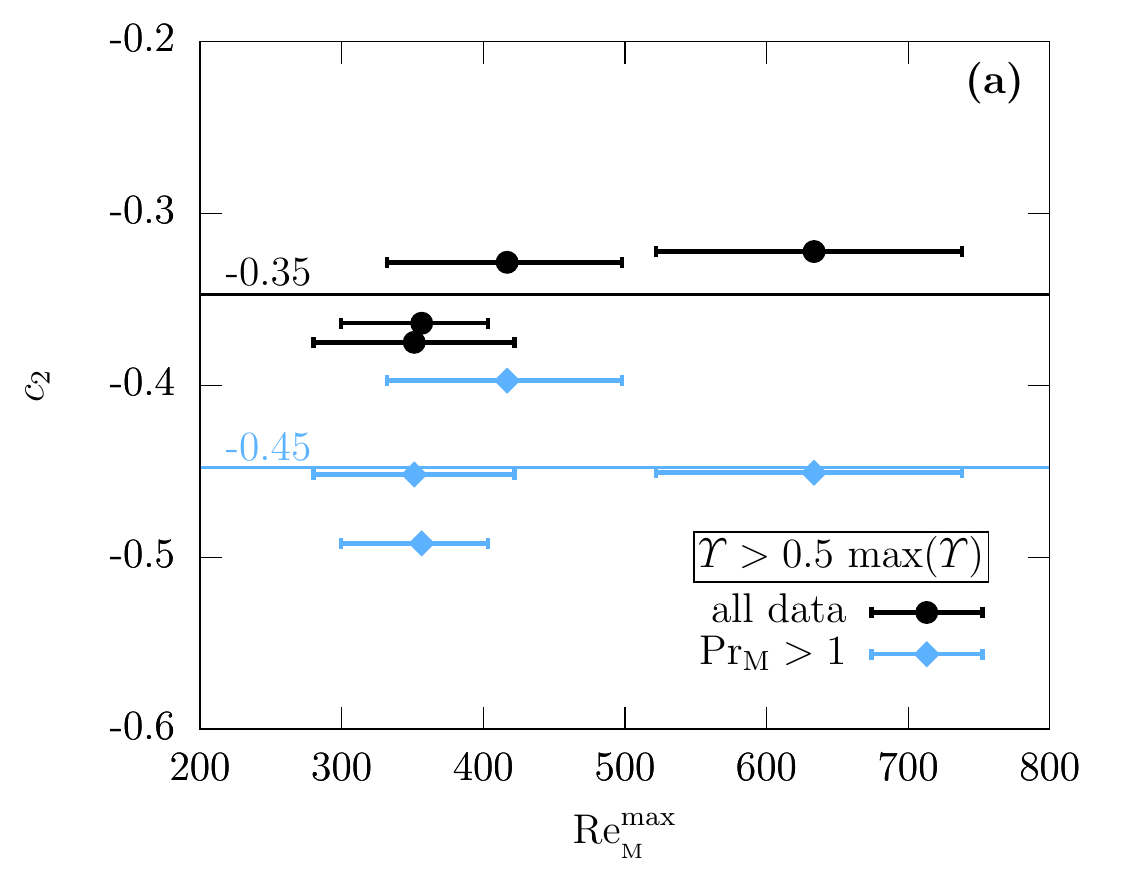}
  \includegraphics[width=0.49\textwidth]{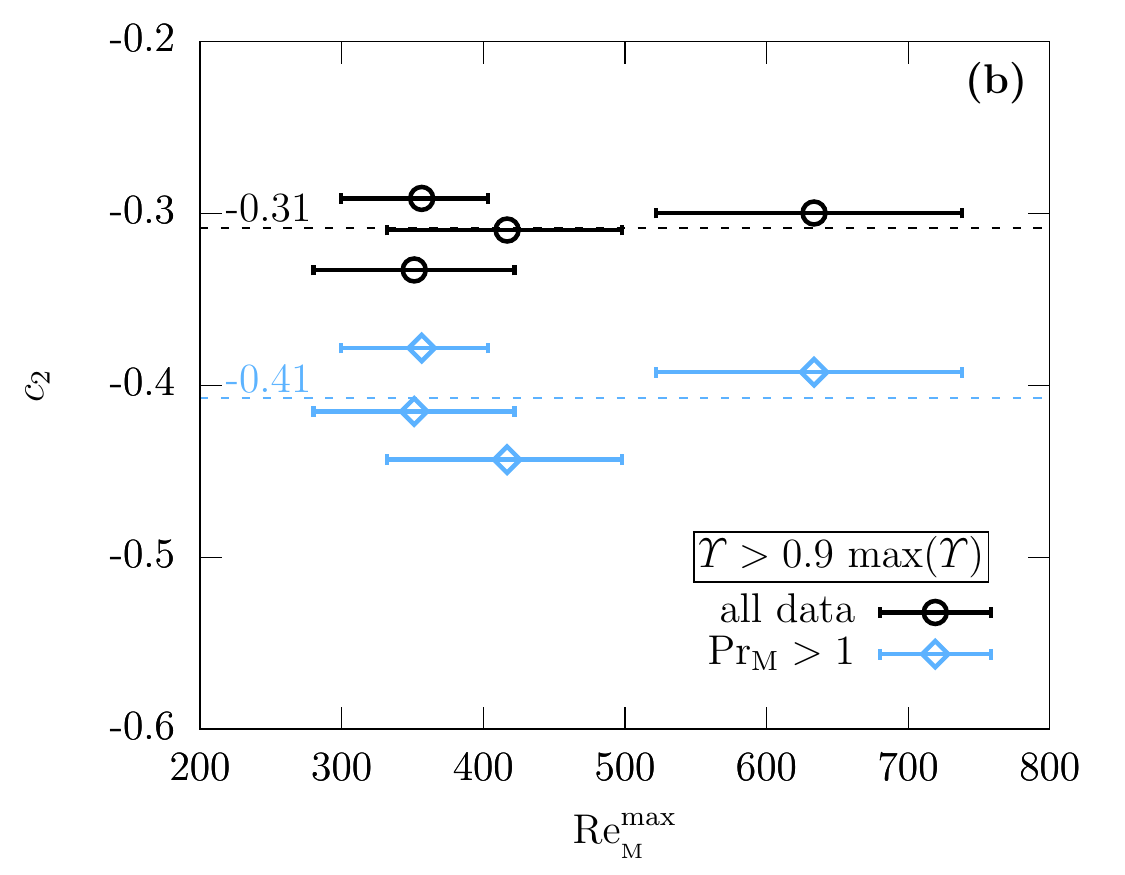}
\caption{
The slope resulting from a fit of the function
$\UpsilonTimeAve = c_1 \Pm^{c_2}$ to the data points
presented in figure~\ref{fig_urmsBrms_Pm__mean}(a) as a function
of the range of maximum magnetic Reynolds numbers found in simulations
shown in figure~\ref{fig_urmsBrms_Pm__mean}(b).
The horizontal bars indicate the range of $\Rm^\mathrm{max}$, which
decreases from small to large $\Pm$.
Results for the fitting to all data are shown as black symbols and results
for using $\Pm>1$ as blue diamonds.
Horizontal lines indicate the average values of the slopes.
(a)
$\UpsilonTimeAve$ has been obtained using all data with
$\Upsilon > 0.5~\mathrm{max}(\Upsilon)$.
(b)
$\UpsilonTimeAve$ has been obtained using all data with
$\Upsilon > 0.9~\mathrm{max}(\Upsilon)$. (colour online)
}
\label{fig_slope_Rm}
\end{center}
\end{figure}

We now check if a power-law fit, $\UpsilonTimeAve = c_1 \Pm^{c_2}$, is
consistent with the data of the fit parameters $c_1$ and $c_2$.
Both averaging conditions, using all data points for which
$\Upsilon > 0.5~\mathrm{max}(\Upsilon)$ and
$\Upsilon > 0.9~\mathrm{max}(\Upsilon)$,
are considered.
The results for the full range of $\Pm$ as well as for $\Pm>1$ can be
found in the appendix in table~\ref{table_fits}.
Additionally, we present the slopes $c_2$ as a function of the corresponding
range of $\mathrm{Re}_\mathrm{M,max}$ in figure~\ref{fig_slope_Rm}.
Fit results to the $\Upsilon > 0.5~\mathrm{max}(\Upsilon)$
condition are shown in figure~\ref{fig_slope_Rm}(a)
and the case of $\Upsilon > 0.9~\mathrm{max}(\Upsilon)$ in figure~\ref{fig_slope_Rm}(b).
The obtained value of $c_2$ is presented for fitting to all available data as
black symbols and for $\Pm>1$ as blue ones.
We do not find a clear dependence of $c_2$ on the Reynolds number range.
When using data for the full $\Pm$ regime explored in this paper, we find mean
slopes between $c_2=-0.31$ and $-0.35$.
The slope of the function $\UpsilonTimeAve(\Pm)$ becomes steeper with values
between $c_2=-0.41$ and $c_2=-0.45$, when fitting only to data with $\Pm>1$.
The latter should be a better description for the large Prandtl number regime,
since the scaling of $\UpsilonTimeAve$
might change in the transition from $\Pm<1$ to $\Pm>1$.

\section{Chiral magnetically driven turbulence in the early Universe}
\label{sec_EU}

The findings from DNS presented above can be leveraged to estimate the
turbulent velocities and the
Reynolds number in the early Universe.
As we have seen in the previous section, the ratio of kinetic to
magnetic energy depends on the magnetic Prandtl number.
Hence, as a first step we estimate the value of $\Pm$ in the early Universe.
Afterwards we estimate $u_\mathrm{rms}$ and the magnetic Reynolds number
for chiral magnetically driven turbulence.

\subsection{Magnetic Prandtl number}

The magnetic Prandtl number has been defined before as the ratio of viscosity
over magnetic diffusivity.
Hence it measures the relative strength of these two transport coefficients.
The derivation of transport coefficients in weakly
coupled high temperature gauge theories has been presented in
\citet{ArnoldEtAl2000} for various matter field content.

For the electric conductivity, \citet{ArnoldEtAl2000} found
a leading term (converted from natural to cgs units)
\begin{equation}
   \sigma_\mathrm{el} =  \frac{\kappa_\sigma}{4\pi\, \alphaem \log\left((4\pi \alphaem)^{-1/2}\right)}
\frac{\kB T}{\hbar}
\end{equation}
with $\kappa_\sigma = 11.9719$ for the largest number of species considered.
The magnetic resistivity in the early Universe follows as
\begin{equation}
  \eta(T) = \frac{c^2}{4\pi\, \sigma_\mathrm{el} }
          = \frac{\alphaem}{\kappa_\sigma} \log\left((4\pi \alphaem)^{-1/2}\right) \frac{\hbar c^2}{\kB T}
          \approx 7.3\times 10^{-4} \frac{\hbar c^2}{\kB T}
          \approx {4.3\times10^{-9}}~T_{100}^{-1}\cm^2\s^{-1}.
  \label{eq_etaEU}
\end{equation}
Here, $T_{100} \equiv 1.2\times10^{15}\K$, so that $\kB T_{100} = 100 \GeV$.

For the shear dynamic viscosity, $\tilde \nu_\mathrm{shear}$,
\citet{ArnoldEtAl2000} report
\begin{equation}
    \tilde  \nu_\mathrm{shear} =
     \frac{\kappa_\mathrm{shear}}{\alphaem^2 \log\left(\alphaem^{-1}\right)} \frac{(\kB T)^3}{\hbar^2 c^3}
\label{eq_nushear}
\end{equation}
with $\kappa_\mathrm{shear}\approx 147.627$ for the largest number of species
considered.
The kinematic viscosity is determined by
$\nu =\tilde \nu_\mathrm{shear}/\meanrho$ with
the mean density in the early Universe being
\begin{equation}
   \meanrho =        \frac{\pi^2}{30}\,g_*\frac{(\kB T)^4}{\hbar^3c^5}
            \approx  7.6\times10^{26}g_{100}T_{100}^4\g\cm^{-3},
\label{eq_meanrho}
\end{equation}
where $g_{100}\equiv g_*/100$ and $g_*=106.75$ is the effective number of
degrees of freedom at $T \approx 100\GeV$ in the Standard Model.
Dividing equation~(\ref{eq_nushear}) by (\ref{eq_meanrho}) we find
the kinematic viscosity
\begin{equation}
   \nu = \frac{30 \kappa_\mathrm{shear}}{\pi^2 g_* \alphaem^2 \log\left(\alphaem^{-1}\right)} \frac{\hbar c^2}{\kB T}
       \approx  1.6\times10^{4} \frac{\hbar c^2}{\kB T}
       \approx  {9.4\times10^{-2}}~T_{100}^{-1}\cm^2\s^{-1} .
\label{eq_nuEU}
\end{equation}

The ratio of equations~(\ref{eq_nuEU}) and (\ref{eq_etaEU}) yields the
magnetic Prandtl number
\begin{equation}
   \Pm = \frac{\nu}{\eta}
       \approx 2.2\times10^{7}.
\label{eq_PmEU}
\end{equation}
We should clarify in this context that this is the
microphysical magnetic Prandtl number and not the turbulent one,
which is always of the order of unity \citep{KR94,YBR03,Jur11}
and independent of the physical conditions.

\subsection{Magnetic Reynolds number}

Assuming that the kinetic energy reaches a fraction $\Upsilon(\Pm)$ of the
magnetic energy at dynamo saturation, it is customary to estimate in physical units
\begin{equation}
   u_\mathrm{rms} \approx \left(\Upsilon(\Pm) \frac{B_\mathrm{sat}}{4\pi \meanrho}\right)^{1/2}.
\label{eq_uEU1}
\end{equation}
The mean density in the early Universe is given in equation~(\ref{eq_meanrho})
and the value of $B_\mathrm{sat}$ depends on the chiral nonlinearity parameter
\begin{equation}
  \lambda=3 \hbar c\,\left({8\alphaem\over\kB T}\right)^2\approx1.3\times10^{-17}\,T_{100}^{-2}\,\cm\erg^{-1}
  \label{eq_lambda_1}
\end{equation}
and the initial chiral chemical potential $\mu_0$.
Since the latter is unknown, we estimate it via the
thermal energy density:
\begin{equation}
  \mu_0 = \vartheta~4\alphaem~\frac{\kB T}{\hbar c}
    \approx 1.5\times10^{14}~ \vartheta~ T_{100} \cm^{-1}.
\label{eq_mu0_EU}
\end{equation}
Due to the uncertainties in $\mu_0$ we introduce the free parameter $\vartheta$,
allowing us to explore different initial conditions.
The magnetic field produced by chiral dynamos as discussed in this paper reaches
a maximum value of
\begin{equation}
   B_\mathrm{sat} = \left(4\pi \frac{\mu_0 k_\lambda}{\lambda}\right)^{1/2}
     \approx 6.2\times10^{21}~ \vartheta~ T_{100}^2 \G,
\label{}
\end{equation}
where we use equation~(\ref{klambda}) for the inverse magnetic correlation length,
that results in
\begin{equation}
   k_\lambda \approx 2.6\times10^{11}~ \vartheta~ T_{100} \cm^{-1}.
\label{eq_klamEU}
\end{equation}
Following equation~(\ref{eq_uEU1}), the magnetic field drives turbulence with an
rms velocity of
\begin{equation}
  u_\mathrm{rms} =6.1\times 10^7 \Upsilon(\Pm)^{1/2} \vartheta \cm \s^{-1}.
  \label{eq_uEU}
\end{equation}

Finally, using equations~(\ref{eq_uEU}), (\ref{eq_klamEU}), and (\ref{eq_etaEU}),
we find the following value for the magnetic
Reynolds number in the early Universe:
\begin{equation}
  \Rm \approx \frac{u_\mathrm{rms}}{k_\lambda\eta}
      \approx 5.5\times 10^4 ~\Upsilon(\Pm)^{1/2}.
  \label{eq_RmEU}
\end{equation}
Note, that the magnetic Reynolds number is based on the
wavenumber $k_\lambda$ which determines the maximum scale of turbulent motions.
We also stress that the size of the inertial range is independent of $\vartheta$,
and hence of $\mu_0$. This is because both,
the forcing scale $k_\lambda$ and the
initial energy input scale $k_\mu$, scale linear with $\mu_0$.
Combining the estimate given by equation~(\ref{eq_RmEU}) together with the
the extrapolation of $\Upsilon(\Pm) \approx 0.08\, \Pm^{-0.4}$ found in DNS,
see section~\ref{sec_PmDependence}, we find $\Rm = \mathrm{O}(10^3)$
when using $\Pm = \mathrm{O}(10^7)$.

\section{Conclusions}

In this paper we analyse the energetics of the chiral magnetically driven
turbulence.
This type of turbulence is produced by the chiral dynamo instability
that originates from an asymmetry between left- and right-handed fermions.
This magnetic field instability
is formally similar to the classical $\alpha^2$ dynamo.
However, while the classical $\alpha^2$ dynamo requires
an energy input by turbulence,
the chiral dynamo creates turbulence via the Lorentz force.
By solving the set of chiral MHD equations in numerical simulations,
we explore the dependence of chiral magnetically driven
turbulence on initial chiral asymmetries and the magnetic Prandtl number.

Our main findings from DNS may be summarized as follows:
\begin{itemize}
\item{For a large range of parameters, it has been shown that the chiral
magnetic instability generates turbulence.
In this paper we have focused on the case of small chiral nonlinearity
parameters $\lambda_\mu$, defined in equation~(\ref{eq_lambdamu}), where turbulence
becomes strong enough to affect the evolution of the magnetic field.
}
\item{The transfer of energy from the chiral chemical potential via
magnetic energy to kinetic energy has been analysed in DNS.
In particular, we found that the ratio of the production rate of kinetic
energy over the production rate of magnetic energy, $\Phi$, increases
exponentially in time during the chiral dynamo phase.
At dynamo saturation, $\Phi$ appears to approach unity, when the magnetic correlation
length remains inside the numerical domain.
} 
\item{A central parameter explored in our simulations is $\Upsilon$,
the ratio of kinetic over magnetic energy; see definition in
equation~(\ref{eq_Upsilon}).
Due to the Lorentz force, the velocity field grows at a
rate that is twice the one of the magnetic field strength.
As a result, $\Upsilon$ increases initially exponentially.
Once there is a back reaction of the velocity field on the magnetic field,
$\Upsilon$ stays approximately constant, see e.g.\ 
figure~\ref{fig_urmsBrms_t}(a).
}
\item{For magnetic Prandtl numbers $\Pm=1$,
the time average of $\Upsilon$, taken after its exponential growth phase
and referred to here as $\UpsilonTimeAve$,
has been determined to be between $0.06$ and $0.07$.
This value seems to be independent of the initial chiral asymmetry.
} 
\item{For $\Pm>1$, the parameter $\Upsilon$ decreases.
With our DNS we find a scaling of approximately
$\Upsilon(\Pm) = 0.1\, \Pm^{-0.4}$.} 
\item{We do not find a change of the function $\Upsilon(\Pm)$ for different
regimes of $\Rm$, however, only a small variation of $\Rm$ has been considered
and this might change when increasing the statistics and the extending the
range of $\Rm$.}
\end{itemize}

A chiral dynamo instability and hence chiral magnetically driven turbulence
can only occur in extreme astrophysical environments, because a high
temperature is required for the existence of a chiral asymmetry.
At low energies chiral flipping reactions destroy any difference in number
density between left- and right handed fermions.
As an astrophysically relevant regime, we have discussed the plasma of the
early Universe.
Our findings from DNS allow to estimate the magnetic Reynolds number
in the early Universe.
In particular, a value of
$\Rm = \mathrm{O}(10^3)$ can be expected for
chiral magnetically driven turbulence, if the chiral asymmetry is generated
by thermal processes.

\section*{Acknowledgements}
J.S.\ thanks the organizers of the ``2nd Conference on Natural Dynamos'' for
the interesting and fruitful conference held in Valtice.
We are grateful to the three anonymous referees who helped improving this work
with their suggestions. 
This project has received funding from the
European Union's Horizon 2020 research and
innovation program under the Marie Sk{\l}odowska-Curie grant
No.\ 665667 (``EPFL fellows'').
I.R.\ acknowledges the hospitality of NORDITA, the University of Colorado,
the Kavli Institute for Theoretical Physics in Santa Barbara
and the \'Ecole Polytechnique F\'ed\'erale de Lausanne.
We thank for support by the \'Ecole polytechnique f\'ed\'erale de Lausanne, Nordita,
and the University of Colorado through
the George Ellery Hale visiting faculty appointment.
Support through the National Science Foundation
Astrophysics and Astronomy Grant Program (grant 1615100),
the Research Council of Norway (FRINATEK grant 231444),
and the European Research Council (grant number 694896) are
gratefully acknowledged.
Simulations presented in this work have been performed
with computing resources
provided by the Swedish National Allocations Committee at the Center for
Parallel Computers at the Royal Institute of Technology in Stockholm.

\newpage

\appendix

\section{Table of fit results}

\begin{table}[h!]
\centering
\caption{Fit results for $\UpsilonTimeAve = c_1 \Pm^{c_2}$.}
     \begin{tabular}{l|ll|ll}
      \hline
      \hline
	Series & $c_1$ (all $\Pm$)	& $c_2$ (all $\Pm$) & $c_1$ ($\Pm>1$)&  $c_2$ ($\Pm>1$)\\	\hline
D (using $\Upsilon >0.5~\mathrm{max}(\Upsilon)$) &0.06&$-0.33$&0.07&$-0.40$\\
D (using $\Upsilon >0.9~\mathrm{max}(\Upsilon)$) &0.07&$-0.31$&0.09&$-0.44$\\
\hline
F (using $\Upsilon >0.5~\mathrm{max}(\Upsilon)$) &0.07&$-0.37$&0.07&$-0.45$\\
F (using $\Upsilon >0.9~\mathrm{max}(\Upsilon)$) &0.08&$-0.33$&0.09&$-0.42$\\
\hline
G (using $\Upsilon >0.5~\mathrm{max}(\Upsilon)$) &0.06&$-0.36$&0.08&$-0.49$\\
G (using $\Upsilon >0.9~\mathrm{max}(\Upsilon)$) &0.07&$-0.29$&0.08&$-0.38$\\
\hline
H (using $\Upsilon >0.5~\mathrm{max}(\Upsilon)$) &0.07&$-0.32$&0.08&$-0.45$\\
H (using $\Upsilon >0.9~\mathrm{max}(\Upsilon)$) &0.07&$-0.30$&0.08&$-0.39$\\
      \hline
      \hline
mean (using $\Upsilon >0.5~\mathrm{max}(\Upsilon)$)&0.07&$-0.35$&0.08&$-0.45$\\
mean (using $\Upsilon >0.9~\mathrm{max}(\Upsilon)$)&0.07&$-0.31$&0.09&$-0.41$\\
      \hline
      \hline
    \end{tabular}
  \label{table_fits}
\end{table}

\end{document}